\documentclass[a4paper,11pt]{article}

\usepackage{jheppub}
\usepackage[T1]{fontenc}
\usepackage[utf8]{inputenc}
\usepackage{mathrsfs} 
\usepackage{multirow}
\usepackage{slashed}
\usepackage{color}
\usepackage{xcolor}
\usepackage{graphicx,subcaption}
\usepackage{cleveref}
\usepackage{enumerate}

\newcommand\MeV{\ \mathrm{MeV}}
\newcommand\GeV{\ \mathrm{GeV}}

\newcommand\ifb{\ {\mathrm{fb}^{-1}}}

\newcommand\xSigmaSMmath{\Sigma x \mathrm{SM}}
\newcommand\xSigmaSM{$\xSigmaSMmath$~}
\newcommand\mdelta{{\Delta m}}
\newcommand\muH{{\mu_H^2}} 
\newcommand\muSig{{\mu_{\Sigma}^2}} 
\newcommand\muS{{\mu_{S}^2}} 
\newcommand\lamSig{{\lambda_\Sigma}}
\newcommand\lamH{{\lambda_H}} 
\newcommand\lamS{{\lambda_S}}
\newcommand\cubS{{a_{ S}}}
\newcommand\tadS{{b}}
\newcommand\cubHSig{{a_{H \Sigma}}}
\newcommand\cubHS{{a_{H S}}}
\newcommand\cubSigS{{a_{\Sigma S}}}
\newcommand\lamHSig{{\lambda_{H \Sigma}}}
\newcommand\lamHS{{\lambda_{H S}}}
\newcommand\lamSigS{{\lambda_{\Sigma S}}}
\newcommand\lamHSigS{{\lambda_{H \Sigma S}}} 

\newcommand\mrm[1]{\mathrm{#1}}

\newcommand\Ztwo{$\mathbb{Z}_2$~}

\newcommand\Sprime{ {S'} }
\newcommand\Sigprime{ {{\Sigma^0}'} }
\newcommand\generalSigprime{ {{\Sigma}'} } 
\newcommand\mixAngle{ {\theta_{S} } }

\title{A Real Triplet-Singlet Extended Standard Model: Dark Matter
and Collider Phenomenology}

\author[a]{Nicole~F.~Bell,}
\author[a]{Matthew~J.~Dolan,}
\author[a,1]{Leon~S.~Friedrich,\note{Corresponding author.}}
\author[b,c,d]{Michael~J.~Ramsey-Musolf,}
\author[a]{and Raymond~R.~Volkas}
\affiliation[a]{ARC Centre of Excellence for Dark Matter Particle Physics, \\ 
School of Physics, The University of Melbourne,\\ Victoria 3010, Australia}
\affiliation[b]{Tsung-Dao Lee Institute and  School of Physics and Astronomy,\\ Shanghai Jiao Tong University,\\ 800 Dongchuan Road, Shanghai, 200240, China}
\affiliation[c]{Amherst Center for Fundamental Interactions,\\ Department of
  Physics, University of Massachusetts Amherst,\\ Amherst, MA 01003, U.S.A.}
\affiliation[d]{Kellogg Radiation Laboratory, California Institute of Technology,\\ Pasadena, CA 91125, U.S.A.}
\emailAdd{n.bell@unimelb.edu.au} \emailAdd{matthew.dolan@unimelb.edu.au}
\emailAdd{lfriedrich@student.unimelb.edu.au}
\emailAdd{mjrm@physics.umass.edu}
\emailAdd{raymondv@unimelb.edu.au}

\begin{document}

\hfill{ACFI T20-12}

\abstract{
We examine the collider and dark matter phenomenology of the Standard Model extended by a hypercharge-zero SU(2) triplet scalar and gauge singlet scalar. In particular, we study the scenario where the singlet and triplet are both charged under a single \Ztwo symmetry.
We find that such an extension is capable of generating the observed dark matter density, while also modifying the collider phenomenology such that the lower bound on the mass of the triplet is smaller than in minimal triplet scalar extensions to the Standard Model.
A high triplet mass is in tension with the parameter space that leads to novel electroweak phase transitions in the early universe. Therefore, the lower triplet masses that are permitted in this extended model are of particular importance for the prospects of successful electroweak baryogenesis and the generation of gravitational waves from early universe phase transitions.   
   }

\maketitle

\section{Introduction}
\label{sec:intro}

The prospect of a strongly first order electroweak phase transition (EWPT) is of great interest due to its potential to generate the
observed baryon asymmetry via electroweak baryogenesis (EWBG), and the prospect of
such a transition resulting in a detectable gravitational wave background. For reviews on EWBG, phase transition induced gravitational waves, and the EWPT in general, see refs.~\cite{TroddenReview,MorrisseyMJRMReview}, \cite{Weir:2017wfa,GrahamReview}, and \cite{GrahamReview,Ramsey-Musolf:2019lsf}, respectively. However, non-perturbative lattice studies show that instead of featuring a first order transition, the Standard Model (SM) EWPT proceeds via a crossover transition~\cite{Kajantie:1996qd,Kajantie:1996mn,Gurtler:1997hr,Laine:1998jb,Csikor:1998eu,Aoki:1999fi}. 
Therefore, the SM can neither generate the observed baryon asymmetry, nor produce detectable gravitational waves during the EWPT.

However, if there is new physics that couples to the SM Higgs boson, the phase transition may instead be first order.
This is often achieved by introducing
gauge singlet scalars~\cite{Espinosa:1993bs,Benson:1993qx,Choi:1993cv,Vergara:1996ub,Ham:2004cf,Ahriche:2007jp,Profumo:2007wc,Noble:2007kk,Espinosa:2007qk,Espinosa:2008kw,Barger:2007im,Ashoorioon:2009nf,Das:2009ue,Espinosa:2011ax,Cline:2012hg,Chung:2012vg,Barger:2011vm,Huang:2012wn,Damgaard:2013kva,Fairbairn:2013uta,No:2013wsa,Profumo:2014opa,Craig:2014lda,Curtin:2014jma,Chen:2014ask,Katz:2014bha,Kozaczuk:2015owa,Kanemura:2015fra,Damgaard:2015con,Huang:2015tdv,Kanemura:2016lkz,Kotwal:2016tex,Brauner:2016fla,Huang:2017jws,Chen:2017qcz,Beniwal:2017eik,Cline:2017qpe,Kurup:2017dzf,Alves:2018jsw,Li:2019tfd,Gould:2019qek,Kozaczuk:2019pet,Carena:2019une,Heinemann:2019trx}. Scalar singlet extensions are capable of generating the desired phase transition and have the benefit of being weakly constrained at colliders due to their small production cross sections.

Another class of models that can feature a strongly first order transition are those that introduce additional scalar SU(2) multiplets. This class of model generally features more complex phenomenology and faces more severe collider constraints.
For example the extensively studied two-Higgs-doublet model~\cite{Turok:1991uc,Davies:1994id,Hammerschmitt:1994fn,Cline:1996mga,Fromme:2006cm,Cline:2011mm,Dorsch:2013wja,Dorsch:2014qja,Harman:2015gif,Basler:2016obg,Dorsch:2017nza,Bernon:2017jgv,Andersen:2017ika,Kainulainen:2019kyp,Zhou:2020xqi}, which is typically inspired by supersymmetry~\cite{Carena:1996wj,Delepine:1996vn,Cline:1996cr,Laine:1998qk,Carena:2008vj,Cohen:2012zza,Laine:2012jy,Curtin:2012aa,Carena:2012np,Katz:2015uja},
and SU(2) triplet scalar extension models~\cite{MJRMTripletPheno,Chowdhury:2011ga,StepInto,MorriseyTwoStep,mjrm_triplet_lattice_1,MJRMMultiplets,TripletPheno,100TeV_Triplet_Pheno,mjrm_triplet_lattice_2} fall into this category.
The addition of a triplet scalar is the simplest scalar SU(2) multiplet extension in the sense that it has the fewest additional physical particles, and the fewest new parameters present without imposing additional symmetries.

SU(2) triplet scalars have been studied extensively in the context of dark matter
(DM) models~\cite{StrumiaMinimalDM,StrumiaSommerfeld,StrumiaCosmic,TripletDM2,MultipletEWPTDM,TripletDM1,LHCTripDM,TripDMFootprint,MJRMMultiplets,100TeV_Triplet_Pheno}, their contribution to the EWPT~\cite{StepInto,mjrm_triplet_lattice_1,TripletGravWaves,mjrm_triplet_lattice_2}, and their collider
phenomenology~\cite{MJRMTripletPheno,LHCTripletPheno2013,planckTriplet,LHCTripletPheno,TripletPheno,100TeV_Triplet_Pheno}.
Unlike singlet scalars, they will always
be produced at colliders via charged and neutral current Drell-Yan processes,
independent of their coupling or mixing with the SM Higgs. 
Reference~\cite{100TeV_Triplet_Pheno} examines the present and future bounds that apply if the neutral component of the triplet is stable. They find that recent disappearing charged track searches using $36 \ifb$ of data require the triplets to have masses larger than about $287\GeV$. If the neutral triplet were unstable, then the lower bound on the mass decreases to around $230\GeV$~\cite{TripletPheno}. Assuming no detection of
new physics, this lower bound will increase with the inclusion of more up-to-date analyses utilising more data. Eventually this lower bound will be in tension
with the parameter space required for a novel EWPT.
Furthermore, while the SU(2) triplet scalar may be stable and contribute to the
DM density, in the
parameter space relevant to a novel EWPT the triplet will only ever contribute a small fraction of the observed DM density.

However, these collider and DM constraints are obviously only applicable in pure triplet scalar extensions. The triplet may simply be one of several additional particles in more complex models. For example, consider real triplet scalars in the Georgi-Machacek model~\cite{Georgi:1985nv,Chanowitz:1985ug,Zhou:2018zli,Zhou:2020idp}, extended supersymmetric models~\cite{THNMSSM}, or those arising from the breaking of some GUT symmetry, e.g., from the {$\bold{210}$} of SO(10)~\cite{SO10SUSYTriplet}. Assuming that at least some of this additional particle content is light, the triplet's collider and DM phenomenology may differ significantly from the minimal model.

In this paper, we investigate how the introduction of a gauge singlet scalar
modifies the collider and DM phenomenology of the SU(2) triplet scalar extended SM. We focus on a subset of models with this particle content where both of the new scalars are charged under a single \Ztwo symmetry and neither scalar gains a vacuum expectation value (VEV) at zero temperature. We find that such an extension is capable of relaxing the lower bound on the mass of the SU(2) triplet, as is desirable for a novel EWPT, while also enabling the production of the correct DM relic density.

The combination of hypercharge-zero SU(2) triplet and gauge singlet scalar extensions has been considered previously in refs.~\cite{THNMSSM,TwoStep,Two_component_SigmaxSM,TripletDM1,TripletDM2}.
Reference~\cite{THNMSSM} examines the phenomenology of a singlet and triplet extended supersymmetric model, where both of the new scalars can gain VEVs at zero temperature. The phenomenology of the model considered in ref.~\cite{THNMSSM} differs significantly from the model considered here as we are not imposing supersymmetry, have only a single Higgs doublet, and none of our scalars gain a zero temperature VEV.
Reference~\cite{TwoStep} considers a singlet-triplet extended two Higgs doublet model in the context of EWBG, but examines neither the DM constraint nor collider constraints due to the production of the new scalars. Once again, the second Higgs doublet prevents a direct analogy. However, some of the collider physics we consider in this work may also be applicable, particularly if the additional Higgs doublet components do not significantly modify the production and decay mechanisms of the singlet and triplet scalars. 
References~\cite{TripletDM2,Two_component_SigmaxSM} examine a model with the same particle content that we consider here. However, they consider a subset of the parameter space where an additional \Ztwo symmetry results in both the singlet and neutral component of the SU(2) triplet being stable and, therefore, contributing to the DM density. As a result, their study of this two-component DM model focuses on a complementary region of parameter-space that features very different collider and DM phenomenology. Reference~\cite{TripletDM1} considers the same model that we examine here, with a focus on the dark matter phenomenology. We extend this previous work with a more detailed discussion of the collider phenomenology, updated direct detection constraints, and the inclusion of indirect detection and electroweak precision constraints. 

The remainder of this paper is structured as follows. In section~\ref{sec:model} we define our model, motivate the study of a specific \Ztwo symmetric sub-model, and  outline our scalar coupling constraints and parameterisation.
We then examine the dark matter and collider phenomenology of our model in sections~\ref{sec:dm}~and~\ref{sec:Collider}, respectively. Finally, we conclude in section~\ref{sec:conclusion}.

\section{Model}
\label{sec:model}

We extend the Standard Model by adding both a
real scalar singlet $S$, and a real scalar field $\Sigma$ transforming
as $(1,3,0)$ under the $SU(3)\times SU(2)\times U(1)_Y$ SM gauge group. We refer to this model as the \xSigmaSM.
We consider the most general renormalisable scalar
potential,

\begin{equation}
\begin{aligned}
  V_{\xSigmaSMmath}
  \ =&
  \ -\ \muH H^\dagger H
  \ -\ \frac{1}{2}\muSig \mathrm{Tr}(\Sigma^2)
  \ -\ \frac{1}{2} \muS S^2
  \\ &
  \ + \ \lamH (H^\dagger H)^2
  \ + \ \frac{1}{4} \lamSig \mathrm{Tr}(\Sigma^2)^2
  \ + \ \frac{1}{4} \lamS S^4
  \\ &
  \ + \ \frac{1}{\sqrt{2}} \cubHSig H^\dagger \Sigma H
  \ + \ \cubHS H^\dagger H S
  \ + \ \frac{1}{2} \cubSigS \mathrm{Tr}(\Sigma^2) S
  \ + \ \frac{1}{3} \cubS S^3
  \\ &
  \ + \ \frac{1}{2} \lamHSig \mathrm{Tr}(\Sigma^2) H^\dagger H
  \ + \ \frac{1}{2} \lamHS  H^\dagger H S^2
  \ + \ \frac{1}{4} \lamSigS  \mathrm{Tr}(\Sigma^2) S^2
  \\ &
  \ + \ \frac{1}{\sqrt{2}} \lamHSigS H^\dagger \Sigma H S
  \ + \ \tadS S
  \,, \label{eq:V0}
\end{aligned}
\end{equation}
where $H$ is the SM Higgs doublet, and we use the notation
\begin{equation}
  \Sigma = \begin{bmatrix}
    \frac{1}{\sqrt{2}} \Sigma^0 & \Sigma^+ \\
    \Sigma^- & -\frac{1}{\sqrt{2}} \Sigma^0
  \end{bmatrix}
  \  , \quad
  H = \begin{bmatrix}
    H^+ \\
    \frac{1}{\sqrt{2}} ( H^0 + i A^0)
  \end{bmatrix} \, .
\end{equation}

This Lagrangian has many parameters and can result in widely different
phenomenology depending on what values the couplings take and whether any are disallowed by additional symmetries. By introducing a \Ztwo symmetry under which just $S$ and $\Sigma$ transform, this model can be categorised into four types based on the \Ztwo charge assignments. The following sections will provide
an overview of the expected phenomenology of each type of model. The remainder
of the paper will focus on the fourth type that we consider, which is outlined last in
section~\ref{sec:typeIV}, as it has the most phenomenological promise from the perspective of achieving EWBG.

\subsection{Type-I, \Ztwo: $S\rightarrow S$, $\Sigma \rightarrow -\Sigma$ }
\label{sec:typeI}

In this scenario, only the triplet is charged under a \Ztwo symmetry. This results in a stable triplet, such that $\Sigma^0$ contributes to the DM density. The singlet will, in general, mix with the
SM Higgs through the $\cubHS$ and $\lamHS$ couplings.

In the absence of the singlet, this scenario corresponds to the pure $\Sigma$ extended SM ($\Sigma$SM) with a stable triplet. As is discussed in refs.~\cite{TripletPheno,100TeV_Triplet_Pheno}, the
parameter space for a light stable triplet is strongly constrained by dark
matter direct detection experiments. However, the introduction of the singlet provides an additional annihilation channel for the triplets in the early
universe, thus reducing their relic density. Furthermore, the singlet can
reduce the direct detection scattering cross section. In the $\Sigma$SM the
DM-nucleus scattering is dominated by a Higgs-exchange diagram, shown in figure~\ref{fig:feyn_dm_scatter_1}, which is now complemented
by a singlet-exchange diagram, shown in figure~\ref{fig:feyn_dm_scatter_2}. There may be regions of the parameter space where these two scattering diagrams partially cancel, thus reducing
the direct detection scattering rate. The combination of these two factors
may open up some of the previously excluded parameter space of triplet
couplings.

\begin{figure}
    \centering
    \begin{subfigure}[b]{0.25\textwidth}
         \centering
         \includegraphics[width=\textwidth]{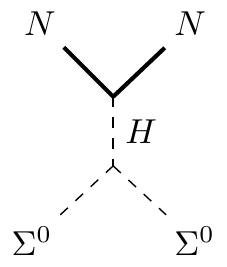}
         \caption{}
         \label{fig:feyn_dm_scatter_1}
     \end{subfigure}
     \hspace{5em}
    \begin{subfigure}[b]{0.25\textwidth}
         \centering
         \includegraphics[width=\textwidth]{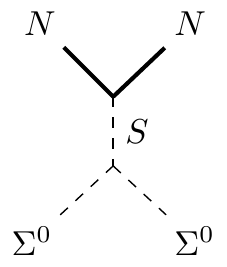}
         \caption{}
         \label{fig:feyn_dm_scatter_2}
     \end{subfigure}
    \caption{DM-nucleus scattering diagrams for the type-I $\xSigmaSMmath$, including; (\subref{fig:feyn_dm_scatter_1}) the SM Higgs mediated scattering present in the $\Sigma$SM, and (\subref{fig:feyn_dm_scatter_2}) a singlet scalar mediated scattering, which is dependent the singlet mixing with the SM Higgs. The nucleus is represented by $N$, the dark matter by $\Sigma^0$.  }
    \label{fig:feyn_dm_scatter}
\end{figure}

The constraints on the singlet scalar couplings will be similar to those
appearing in the real singlet scalar extended standard model, which has already been studied extensively in the context of electroweak phase transitions~\cite{Espinosa:1993bs,Benson:1993qx,Choi:1993cv,Vergara:1996ub,Ham:2004cf,Ahriche:2007jp,Profumo:2007wc,Noble:2007kk,Espinosa:2007qk,Espinosa:2008kw,Barger:2007im,Ashoorioon:2009nf,Das:2009ue,Espinosa:2011ax,Cline:2012hg,Chung:2012vg,Barger:2011vm,Huang:2012wn,Damgaard:2013kva,Fairbairn:2013uta,No:2013wsa,Profumo:2014opa,Craig:2014lda,Curtin:2014jma,Chen:2014ask,Katz:2014bha,Kozaczuk:2015owa,Kanemura:2015fra,Damgaard:2015con,Huang:2015tdv,Kanemura:2016lkz,Kotwal:2016tex,Brauner:2016fla,Huang:2017jws,Chen:2017qcz,Beniwal:2017eik,Cline:2017qpe,Kurup:2017dzf,Alves:2018jsw,Li:2019tfd,Gould:2019qek,Kozaczuk:2019pet,Carena:2019une,Heinemann:2019trx}.
The constraints may differ slightly
due to the possibility of $S\rightarrow \Sigma \Sigma $ decays and new collider
production diagrams involving the triplets. Nonetheless, we expect the singlet to be relatively unconstrained, provided we enforce $m_S \gtrsim m_H/2$ to avoid the decay $H\rightarrow S S $.

However, as discussed in
refs.~\cite{MJRMTripletPheno,TripletPheno,100TeV_Triplet_Pheno}, there is also a
disappearing charged track constraint which currently requires $m_\Sigma\gtrsim
287\GeV$. This constraint is not affected by the singlet scalar, and thus will exclude most of the parameter space of interest to us. Therefore we will not  examine the Type-I \xSigmaSM further in this paper.

\subsection{Type-II, \Ztwo: $S\rightarrow - S$, $\Sigma \rightarrow \Sigma$ }
\label{sec:typeII}

In this \xSigmaSM variant, only the singlet is charged under a \Ztwo symmetry. Thus the singlet will be stable and contribute to the DM density, while the triplet will mix with the SM Higgs due to the $\cubHSig$ coupling, though this mixing must be small due to $\rho$-parameter constraints.

Minimal singlet scalar DM models are strongly constrained by competing
requirements on the singlet's couplings with the SM Higgs. On one hand, this coupling has
to be small as the DM direct detection scattering is mediated by the SM Higgs.
On the other, the coupling has to be large enough to allow the singlet to
annihilate in the early universe, in order to avoid an over-density of dark
matter. There is an allowed region of the singlet scalar parameter
space near $m_S \sim m_H/2$, due to a resonance of the DM annihilation rate, but
otherwise, most of the parameter space is excluded (see refs.~\cite{gambit_xsm_dm_scalar,higgs_portal_dm_review} and references therein).
These constraints may be relaxed significantly by the presence of the triplet,
as it both opens up a new annihilation channel that is not directly tied to the
direct detection scattering rate ($S S \rightarrow \Sigma \Sigma \rightarrow
\mathrm{SM}$), and the triplet may act to reduce the direct detection scattering
rate in a manner analogous to the type-I scenario.

The collider phenomenology of the triplet will be similar to the $\Sigma$SM phenomenology that has already been discussed extensively in refs.~\cite{MJRMTripletPheno,LHCTripletPheno2013,planckTriplet,LHCTripletPheno,TripletPheno}. The main effect of the singlet is the introduction of the $\Sigma^0
\rightarrow S S$ decay channel. Assuming this decay is kinematically allowed, this will lead to events with large missing energy, which some SUSY chargino-neutralino searches are sensitive to. However, this branching ratio can be made arbitrarily small by tuning the scalar
couplings or by having more massive singlets. For intermediate values of the branching
ratio the effect of the new missing energy signal region will be to siphon events away from other signal regions.

A preliminary examination of collider constraints for this \xSigmaSM type was
carried out using the same methodology used in ref.~\cite{TripletPheno}. We
found that the decrease in events in some signal regions was compensated for by
the gain of events in signal regions with large missing energy, leading to no
significant reduction in the lower bound on the triplet mass.

Therefore, while this type of model may be able to satisfy DM
constraints, most of the parameter space of interest to us is still excluded.
Therefore we will not examine this type of \xSigmaSM model  any further.

\subsection{Type-III, \Ztwo: $S\rightarrow S$, $\Sigma \rightarrow \Sigma$ }
\label{sec:typeIII}

In this scenario, there is no \Ztwo symmetry and all couplings are allowed,
leading to a very large number of free parameters. In general, these couplings will lead to
mixing between all of the scalars. The scalar mixing will result in all the new scalars being
unstable, such that there is no DM candidate.

While the singlet--triplet mixing may be very large, the mixing with
the SM Higgs is constrained by the $\rho$-parameter and Higgs coupling measurements. Therefore, the dominant
production mechanism at colliders will likely still be pair production via
charged and neutral current Drell-Yan processes, as in the $\Sigma$SM. However
in addition to decaying into fermion or weak gauge boson pairs, the new scalars
may also decay into one another. For example, assuming $m_{S'}$ is less than
$m_{{\Sigma^\pm}'}$ and $m_{{\Sigma^0}'}$, we have two additional decays,
\begin{itemize}
\item ${\Sigma^0}' \rightarrow S'^{(*)} S'\,$,
\item ${\Sigma^\pm}' \rightarrow {W^\pm}^{(*)} S'\,$,
\end{itemize}
where we have used primed indices to denote mass eigenstates whose primary component is the corresponding unprimed particle.
Each of the scalars produced in such a decay would subsequently decay into
fermion or weak gauge boson pairs. The most constraining current analyses for
the $\Sigma$SM come about from multilepton searches~\cite{TripletPheno}. These longer scalar decay
chains increase the likelihood of pair production events leading to a multilepton final state, which results in more severe constraints. This may be
partly offset by the lower energies of the final state particles in these longer decay chains,  leading to lower efficiency cuts in the relevant analyses.

Without a more in-depth analysis of the collider phenomenology, it is not clear
whether or not this \xSigmaSM model type can reduce the lower bound on the
triplet mass. Consequently, the parameter space available for novel electroweak phase transitions may still be severely restricted. Additionally, the type-III model has no dark matter candidate. For these reasons, together with the complication of a large number
of phenomenologically relevant parameters, we defer examination of  this model to future work,  and will instead focus on the fourth and final type.

\subsection{Type-IV, \Ztwo: $S\rightarrow - S$, $\Sigma \rightarrow -\Sigma$ }
\label{sec:typeIV}
In this model type both the singlet and triplet are charged under a \Ztwo
symmetry. Thus one or both particles will be stable and contribute to the dark matter relic density. This scenario can be further broken up into two sub-categories depending on whether the new scalars are charged under the same or two different \Ztwo symmetries.

If $\lamHSigS = 0$, which effectively corresponds to there being two separate
\Ztwo symmetries, then the neutral component of both the triplet and the singlet
are stable and both will contribute to the DM relic density. It is likely that this scenario will face the same severe constraints that are encountered in minimal triplet DM model with $\muSig>0$. In principle the couplings with the new singlets may
reduce the triplet DM density somewhat via $\Sigma \Sigma \rightarrow S S$
annihilation, and this scenario may be worth examining. However, the constraints
from disappearing charged tracks, as discussed in section~\ref{sec:typeI}, will
still apply. Thus, a low-mass triplet would still be excluded in this scenario.
Note that this sub-type with $\lamHSigS=0$ is the same model as the two-component DM model considered in ref.~\cite{Two_component_SigmaxSM}. However, they allow for $\muSig<0$, and thus allow the triplet component of the DM to avoid direct detection constraints. 

If instead both scalars are charged under the same \Ztwo symmetry, which allows $\lamHSigS \neq 0$, then after electroweak
symmetry breaking the $\lamHSigS$ term will induce mixing between the singlet and neutral component of the triplet. Therefore, if the singlet-like mass eigenstate is lighter than the triplet-like one, then the DM will only consist of singlet-like particles, allowing us  to avoid the severe triplet DM direct detection constraints that are present when $\muSig<0$.

The triplets can still be produced at colliders as in the $\Sigma$SM, via
charged and neutral current Drell-Yan processes. However, unlike the $\Sigma$SM,
the $\lamHSigS$ mixing term  allows the triplets to decay rapidly into the neutral DM
candidate and SM particles. This leads to events with large missing energy and
removes the disappearing charged track constraint.

This model has the prospect of both alleviating collider constraints on triplet scalars, thus increasing the parameter space available for novel electroweak phase transitions, and presenting a viable dark matter candidate. Therefore, it is this version
of the \xSigmaSM that we will study for the remainder of this paper.

\subsubsection{Mass Matrix and Mixing Angles}
\label{sec:mass}

After electroweak symmetry breaking, the $\lamHSigS$ term will lead to a mass
term that will cause mixing between the singlet scalar and neutral component of
the triplet,
\begin{subequations}
\begin{align}
  V_{\xSigmaSMmath} & \supset
  \frac{1}{2}\begin{pmatrix}
    \Sigma^0 & S
  \end{pmatrix}
  \begin{pmatrix}
    - \muSig + \frac{1}{2 } v_H^2 \lamHSig  & - \frac{1}{4} v_H^2 \lamHSigS  \\ -
    \frac{1}{4} v_H^2 \lamHSigS  & - \muS + \frac{1}{2 } v_H^2 \lamHS
 \end{pmatrix}
  \label{eq:massMatrix}
  \begin{pmatrix}
    \Sigma^0  \\ S
  \end{pmatrix},
  \\ & = \frac{1}{2}\begin{pmatrix}
    \Sigprime & \Sprime
  \end{pmatrix}
  \begin{pmatrix}
   m_{\Sigprime}^2 & 0 \\ 0 & m_{\Sprime}^2
  \end{pmatrix}
  \begin{pmatrix}
    \Sigprime \\ \Sprime
  \end{pmatrix},
\end{align}
\end{subequations}
where we have introduced the mass basis,
\begin{align}
  \begin{pmatrix}
    \Sigprime \\ \Sprime
  \end{pmatrix}
  &=
  \begin{pmatrix}
    \cos \mixAngle &\   \sin \mixAngle \\
    - \sin \mixAngle &\  \cos \mixAngle
  \end{pmatrix}
  \begin{pmatrix}
    \Sigma^0 \\ S
  \end{pmatrix}.
\end{align}
We choose to define the scalar mixing angle $\mixAngle$ such that $\sin^2
\mixAngle \leq 0.5$, in order to ensure that the mass eigenstate labelled
$\Sigprime$ always consists primarily of $\Sigma^0$.
As will be discussed in subsequent sections, the collider and DM phenomenology is very sensitive to the neutral scalar mass difference. As it appears frequently in the discussion, we introduce the notation
\begin{equation}
\label{eq:DeltaM}
    \mdelta = m_{\Sigprime} - m_{\Sprime}\,.
\end{equation}
The tree-level mass of the $\Sigma^+$ is simply given by the first diagonal element
of the neutral scalar mixing matrix in eq.~\eqref{eq:massMatrix}. In the absence of mixing, $\lamHSigS=0$, the masses of the charged and neutral components of
the triplet would be degenerate. However, once the SM Higgs breaks the SU(2) symmetry and gives masses to the $W$ and $Z$ bosons, electroweak radiative corrections to the triplet mass lead to a small mass splitting of about $166 \MeV$~\cite{StrumiaMinimalDM}. For this initial study we neglect the effect that the singlet has on the radiative mass correction, which may be significant for large mixing angles, and instead approximate this splitting by setting.
\begin{equation}
  \label{eq:mSigmaC}
  m_{\Sigma^+}  =  \sqrt{- \muSig + \frac{1}{2 } v_H^2 \lamHSig} + 166 \MeV \, .
\end{equation}
In minimal triplet models, this radiative mass splitting always leads to the charged component of the triplet being more massive than the neutral one.
This is not the case in our model as the SU(2) symmetry breaking neutral scalar mixing can raise the mass of the $\Sigprime$. Figure~\ref{fig:mass} illustrates this by showing how the masses of the scalars behave as a function of $\muS$ for some benchmark parameters. The small range for $\muS$ and small value for $\lamHSigS$ in figure~\ref{fig:mass} were chosen such that the $166\MeV$ radiative mass splitting is clearly visible. Note that the convention of using $\Sigprime$ to denote the mass eigenstate that consists primarily of $\Sigma^0$ leads to a discontinuity in the labelling of the neutral scalar masses and couplings. This discontinuity in the labelling is clearly visible in figure~\ref{fig:mass}. 

\begin{figure}
  \centering
  \includegraphics{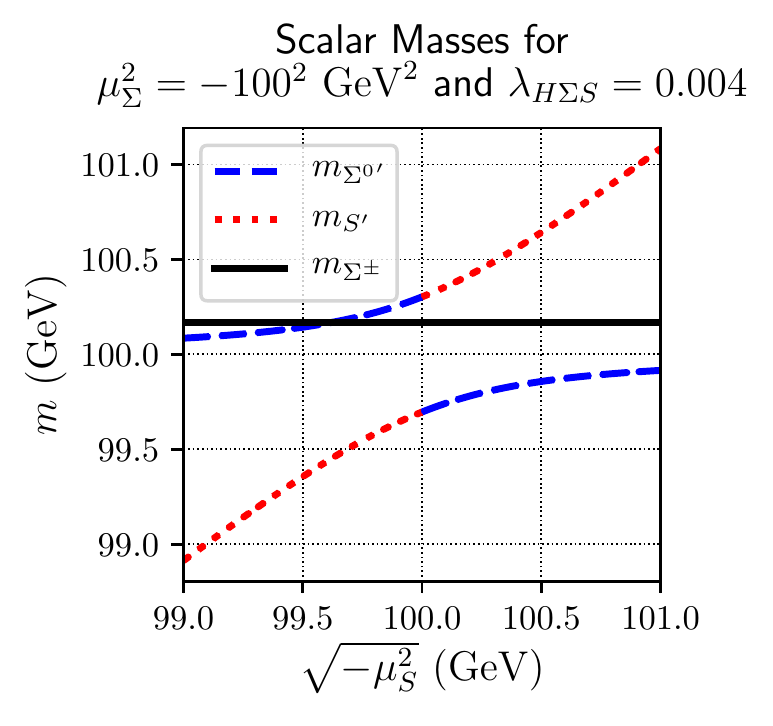}
  \caption{ \label{fig:mass}
        Masses of the new scalars as a function of $\muS$. The solid black, dashed blue, and dotted green lines give the masses of the $\Sigma^\pm$, $\Sigprime$, and $\Sprime$ particles, respectively. All other scalar couplings were set to zero. 
        The identification of $\Sigprime$ as the mass eigenstate that consists primarily of $\Sigma^0$ leads to a discontinuity in the labelling of the mass eigenstates when $\muS = - 100^2 {\mathrm{GeV}}^2$. 
        }
\end{figure}

\subsubsection{Oblique Corrections}
\label{sec:ewpo}

The singlet and triplet scalars will result in new electroweak radiative corrections, which are constrained by electroweak precision data. These corrections can be parameterised by the oblique parameters. The current limits on these parameters are~\cite{pdg2020}
\begin{subequations}
\label{eq:STU_values}
\begin{align}
    \Delta S &= S - S_\mrm{SM} =  - 0.01 \pm 0.10 \, ,
    \\
    \Delta T &= T - T_\mrm{SM} = 0.03 \pm 0.12 \, ,
    \\
    \Delta U &= U - U_\mrm{SM} = 0.02 \pm 0.11 \, .
\end{align}
\end{subequations}

Our new particle content will contribute to the $T$ and $U$ parameters via one-loop diagrams. There is no contribution to the $S$ parameter as we have introduced neither new particles with hypercharge nor new particles that mix with the SM Higgs.
The relevant loop diagrams and resulting loop functions also appear in the minimal SU(2) triplet extension. This allows us to write our contributions as,
\begin{subequations}
\begin{align}
    T_{\mathrm{\xSigmaSMmath}} &= \cos^2{\mixAngle}\, T_{\mathrm{\Sigma SM}}(m_{\Sigprime}) \ + \  \sin^2{\mixAngle}\, T_{\mathrm{\Sigma SM}}(m_{\Sprime}),\\
    U_{\mathrm{\xSigmaSMmath}} &= \cos^2{\mixAngle}\, U_{\mathrm{\Sigma SM}}(m_{\Sigprime}) \ + \  \sin^2{\mixAngle}\, U_{\mathrm{\Sigma SM}}( m_{\Sprime}),
\end{align}
\end{subequations}
where $T_{\mathrm{\Sigma SM}}$, $U_{\mathrm{\Sigma SM}}$ are the leading order one-loop contributions in the minimal triplet extension.\footnote{The minimal triplet model with a non-zero triplet VEV and triplet-Higgs mixing will have tree level and additional one-loop contributions that are not present in our model.} 
These functions can be found in refs.~\cite{Forshaw:2001xq,MJRMTripletPheno,planckTriplet}, and are well approximated by,
\begin{subequations}
\begin{align}
    T_{\mathrm{\Sigma SM}}(m) &\approx \frac{1}{6 \pi} \frac{1}{s_W^2 c_W^2} \frac{( m -  m_{\Sigma^+} )^2}{m_Z^2},\\
    U_{\mathrm{\Sigma SM}}(m) &\approx \frac{m - m_{\Sigma^+}}{3 \pi m_{\Sigma^+}}.
\end{align}
\end{subequations}
These corrections can be significant if the $\lamHSigS$ coupling is large, such that there is a large mixing angle and triplet-component mass differences $(m_{\Sigprime} - m_{\Sigma^+})$.

In order to quantify the constraints coming from electroweak precision observables, we take the same approach used in refs.~\cite{Profumo:2007wc,Profumo:2014opa,Kotwal:2016tex,Li:2019tfd} and define
\begin{subequations}
\begin{align}
    \Delta \chi^2 &= \sum_{i,j} \left(\mathcal{O}_{i,\mathrm{\xSigmaSMmath}} - \Delta \mathcal{O}_i^0\right)
    \left(\sigma^2\right)^{-1}_{i j}
   \left(\mathcal{O}_{j,\mathrm{\xSigmaSMmath}} - \Delta \mathcal{O}_j^0\right) \, , \\
    \sigma^2_{i j}  &= \sigma_i \, \rho_{i j} \, \sigma_j \, ,
\end{align}
\end{subequations}
where $\mathcal{O}_i \in \{S,T,U\}$, $\Delta \mathcal{O}_i^0$ and $\sigma_i$ denote the central values and errors in eqs.~\eqref{eq:STU_values}, and the correlation matrix is~\cite{pdg2020},
\begin{equation}
   \rho_{i j} = \left(
\begin{array}{ccc}
 1 & 0.92 & -0.8 \\
 0.92 & 1 & -0.93 \\
 -0.8 & -0.93 & 1 \\
\end{array}
\right) \, .
\end{equation}
We then consider any points with $\Delta \chi^2 > 7.82$, which corresponds to the $95\%$ C.L.~for three degrees of freedom, to be excluded by electroweak precision observables.

\subsubsection{Parameter Selection and Coupling Constraints}
\label{sec:paramconstraints}
Aside from the ordinary SM Higgs couplings, which are fixed by requiring $m_H=125\GeV$ and $v_H=246\GeV$, there are eight free scalar potential parameters in eq.~\eqref{eq:V0}. However, given that the scalar masses are more phenomenologically relevant, we will instead parameterise our model in terms of the following eight parameters; $m_\Sprime$, $m_\Sigprime$, $\lamSigS$, $\muSig$, $\lamHS$, $\lamSig$, $\lamS$ and $ \lamHSigS$. The first five of these parameters fix all of the components of the mass
matrix, eq.~\eqref{eq:massMatrix}, and thus uniquely determine $\muS$, $\lamHSig$, $m_{\Sigma^\pm}$, and $\mixAngle$. However, note that the eight parameters are not entirely independent, as some combinations will be
unphysical. This can be seen by noting that $m_{\Sigprime}=m_{\Sprime}$
requires $\lamHSigS = 0$, such that a non-zero selection for $\lamHSigS$ would be unphysical. Particularly, physical choices must satisfy,
\begin{equation}
\label{eq:min_mass_diff}
  \left \lvert m_{\Sigprime}^2 - m_{\Sprime}^2 \right \rvert \geq \frac{1}{2} v_H^2 \lamHSigS \, . 
\end{equation}

We also require that the scalar potential is bounded from below. In the absence of
the $\lamHSigS$ coupling, this requirement can be expressed as a simple set of inequalities.
However, with non-zero $\lamHSigS$, the full set of conditions are quite complicated. Vacuum stability conditions are derived in ref.~\cite{bounded_general_potential} for several different models, including one with two real scalar singlets and a SM Higgs doublet. The constraints that apply to that model are also applicable to our model. Additionally, ref.~\cite{bounded_general_potential} conveniently provides a supplementary  Mathematica notebook that includes the necessary inequalities. Therefore, we will simply require that our couplings satisfy the relevant inequalities, eq.~(61) in ref.~\cite{bounded_general_potential}, but we will not replicate them here.

We also require that our scalar couplings are perturbative. This requirement is generally more strict
than simply requiring the couplings satisfy perturbative unitarity (see,
e.g., refs.~\cite{mjrmRunningCouplingsSinglets,mjrmRunningCouplings,TripletPheno}).
Given that $\lamHSig$ is directly related to $m_{\Sigma^+}$ via eq.~\eqref{eq:mSigmaC}, the perturbativity bound on $\lamHSig$ can be translated into a mass bound for $m_{\Sigma^+}$. Therefore, motivated by the bound in ref.~\cite{TripletPheno}, we will ensure $\lamHSig$ is perturbative by requiring $0 < \muSig < 200^2 \ \mathrm{GeV}^2$ and $m_{\Sigma^+}<400\GeV$. For all other quartic scalar couplings $\lambda$, we will simply impose the requirement that $\lambda<2$.

Note that the presence of the triplet will modify the SM Higgs diphoton decay rate. This places constraints on $m_{\Sigma^\pm}$ and $\lamHSig$. Given that this correction has already been discussed extensively in the literature, and that the correction is generally within three standard deviations of the measured value, we will not be discussing this constraint in detail and instead refer readers to refs.~\cite{TripletPheno,LHCTripletPheno,planckTriplet,TwoStep,LHCTripDM}.

\section{Dark Matter Phenomenology}
\label{sec:dm}

In the limit where both $\lamSigS$ and $\lamHSigS$ are small, such that the singlet--triplet interactions are not significant,
the DM phenomenology of our model will be very similar that of a minimal singlet or triplet scalar DM model. Thus, we may have to address the same issues faced by these models, which must somehow be resolved by the introduction of singlet--triplet interactions. In particular, if $m_\Sigprime < m_\Sprime$, such that the triplet-like neutral scalar $\Sigprime$ is the DM candidate, one might expect to encounter the same issues that arise in minimal triplet scalar DM models:
\begin{itemize}
\item The triplet can rapidly annihilate into weak gauge bosons, such that the relic density of light triplets will only ever be a small fraction of the observed DM density.
\item DM direct detection constraints require $\lamHSig$ to be small, which generally requires $\muSig<0$ and is incompatible with scenarios that feature novel electroweak phase transitions due to a triplet VEV in the early universe.
\end{itemize}
Conversely, if $m_\Sprime < m_\Sigprime$, such that the singlet-like neutral scalar $\Sprime$ is the DM candidate, one might expect to encounter the same issues that arise in minimal singlet scalar DM models:
\begin{itemize}
\item The singlet annihilates through its coupling with the SM Higgs, and thus requires $\lamHS$ to be large enough to avoid an over-density of dark matter.
\item Simultaneously, $\lamHS$ must be small enough to avoid direct detection constraints
\item Satisfying both of these conditions requires $m_S \sim m_H/2$, such that the $S S \rightarrow H$ annihilation rate is resonantly enhanced.
\end{itemize}

The \xSigmaSM may be capable of addressing some of these issues faced by the corresponding minimal DM models. In the case where $\Sigprime$ is the DM candidate, a large neutral scalar mixing angle may reduce its coupling with the SM Higgs and weak gauge bosons. This would result in a both a smaller direct detection cross section and a smaller annihilation rate, and thus also a larger relic density. In the case where $\Sprime$ is the DM candidate, the \xSigmaSM features two new annihilation channels, $\Sprime \Sprime \rightarrow \generalSigprime \generalSigprime \rightarrow \mrm{SM}$ and $\Sprime \Sprime \rightarrow W^+ W^-$. These annihilation channels can keep the $\Sprime$ relic density small without requiring a large coupling with the SM Higgs.

To quantitatively examine the DM phenomenology of the \xSigmaSM we use \texttt{MicrOMEGAS} $5.0.8$~\cite{micromegas} to evaluate the relic density and direct detection scattering cross section. We use the observed dark matter density as measured by the Planck collaboration~\cite{planck2018}, $\Omega_{\mrm{DM}} h^2 = 0.12 $, and compare the cross section obtained by \texttt{MicrOMEGAS} to the
XENON1T~\cite{XENON1T1Y} $90\%$-confidence upper bound on the spin-independent scattering
cross section, after scaling to account for the
fraction of the density of DM that is made up of our DM candidate.
\texttt{MicrOMEGAS} also provides tools for the examination of indirect detection constraints arising from tree-level annihilation processes. We found that inclusion of DM indirect detection constraints arising from these annihilation processes did not lead to any additional constraints. However, \texttt{MicrOMEGAS} does not automatically evaluate general loop-induced annihilation processes such as $\Sprime \Sprime \rightarrow h^* \rightarrow \gamma \gamma$. This annihilation into monochromatic photons results in a very clean indirect detection signal. Therefore, we evaluate the 
diphoton annihilation cross section using the analytic formulae given in appendix~\ref{app:indirect} and 
compare this with the current Fermi-LAT limits~\cite{Ackermann:2015lka}.

We explore the \xSigmaSM parameter space by performing a random scan of the scalar potential parameters and evaluating the dark matter density and direct detection constraints for each point. We randomly select dimensionful parameters from a uniform distribution, with masses ranging from $65$ to $300\GeV$ and $\sqrt{\muSig}$ ranging from $0$ to $200\GeV$. We randomly select the dimensionless quartic scalar couplings from a log-uniform distribution ranging from $10^{-4}$ to $2$.
Note that the range of parameters does not include the $m_\Sprime, m_\Sigprime < \frac{m_H}{2}$ region, as that region of parameter space requires one to avoid invisible Higgs decay constraints. This places severe constraints on the $\lamHS$, $\lamHSig$, and $\lamHSigS$ couplings. As one of the motivations for this model is the potential to generate a novel electroweak phase transition, we are interested in regions of parameter space free of these severe constraints on the scalar couplings.

\begin{figure}
  \centering
  {
  \includegraphics{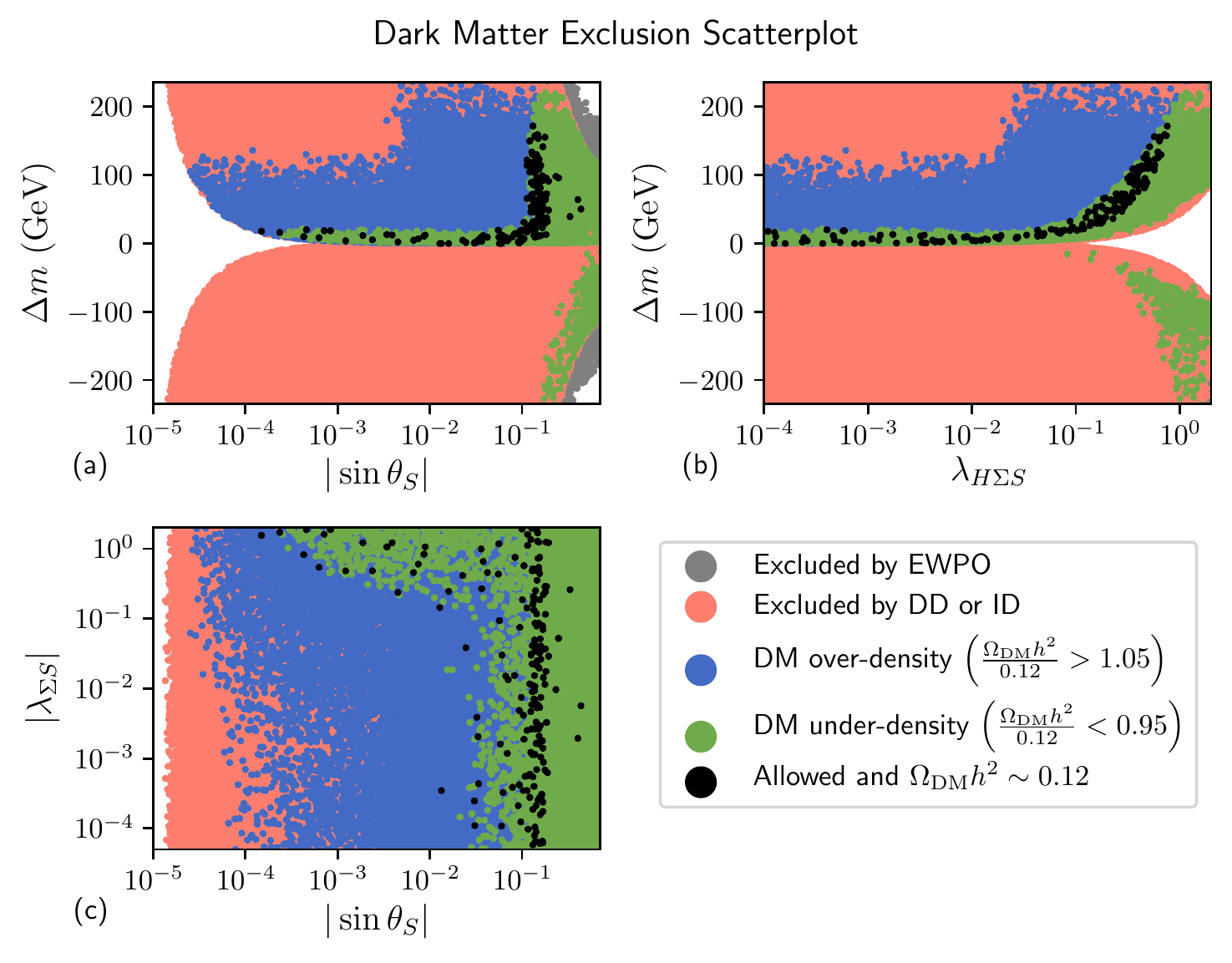}
  \phantomsubcaption\label{fig:DM_scatterplot_a}
  \phantomsubcaption\label{fig:DM_scatterplot_b}
  \phantomsubcaption\label{fig:DM_scatterplot_c}
  }
  \caption{\label{fig:DM_scatterplot}
  Scatter plots showing the dark matter constraints for a random scan of the
  parameter space. Grey points are excluded by electroweak precision observables (EWPO), red points are excluded by DM direct detection (DD) or DM indirect detection (ID), blue points are excluded by
  a DM over-density, green points are allowed but result in a DM under-density, and black points are allowed and approximately yield the correct relic density. 
  As mentioned in the text, the indirect detection constraints are relatively weak and the red points are primarily excluded by direct detection. 
  }
\end{figure}

Figure~\ref{fig:DM_scatterplot} shows a scatter plot of the DM results from the random scan of the parameter space. Grey points are excluded by electroweak precision observables (discussion in section~\ref{sec:ewpo}), red points are excluded by direct or indirect detection, blue points are excluded by DM overproduction, green points are allowed but result in a DM under-density, and black points are allowed and approximately yield the correct relic density.
For the parameter space considered here, the indirect detection constraints are subdominant. While several points are excluded by indirect detection, most of these are also excluded by either direct detection or a DM over-density. Only two points in the scan were excluded solely by indirect detection constraints. 
Examination of figure~\ref{fig:DM_scatterplot} reveals several key features:
\begin{enumerate}
    \item Points with $\mdelta<0$ are excluded by DM direct detection unless $\lamHSigS$, and thus also $\mixAngle$, are large.
    This occurs because when $\mdelta<0$, such that $\Sigprime$ is the DM candidate, our requirement that $\muSig>0$ results in a large $\lamHSig$ and large $\Sigprime$ direct detection cross section. However, the constraint can be avoided  if the $\mixAngle$ mixing angle is large. This allows for cancellation in the $\Sigprime$--$H$ quartic coupling, thus reducing the direct detection cross-section to acceptable levels.
    
    \item The points with $\mdelta<0$ allowed by direct detection do not reproduce the observed relic density.
    This occurs because the $\Sigprime$ coupling to weak gauge bosons tends to result in a small relic density. When combined with the previous point, this means that $\Sigprime$ cannot satisfy both the direct detection and relic density constraints when $\mdelta<0$.
    \item The DM requirements are only satisfied when either $\lvert \lamSigS \rvert \gtrsim 0.1$ or $\lvert \sin \mixAngle \rvert \gtrsim 0.01$ (see black points in figure~\ref{fig:DM_scatterplot_c}). This occurs as the singlet tends to produce an over-density of dark matter, and in order to have the correct relic density it needs a sufficiently large annihilation cross-section. It cannot obtain this through couplings with the SM Higgs\footnote{Except for the region where there is a resonance $m_{\Sprime} \sim m_H/2$ as discussed below.}, as this would be in tension with  direct detection constraints. Therefore it must annihilate either through its couplings with the triplet $\lamSigS$ ($\Sprime \Sprime \rightarrow \Sigprime \Sigprime \rightarrow \mathrm{SM}$), or through its coupling to the weak gauge bosons, which it acquires through its mixing with the triplet, $\propto \sin\mixAngle$. Therefore at least one of $\lamSigS$ and $\mixAngle$ must be large.
    \item When $\lvert\sin \mixAngle \rvert \lesssim 0.1$, the only points that satisfy all DM requirements are those with $0 \lesssim \mdelta \lesssim 30\GeV$ (the thin horizontal strip of black points in figure~\ref{fig:DM_scatterplot_a}). As established in the previous point, obtaining the correct singlet relic density
    when $\lvert\sin \mixAngle \rvert \lesssim 0.1$
    relies on the $\Sprime \Sprime \rightarrow \Sigprime \Sigprime \rightarrow \mathrm{SM}$ annihilation channel. This rate is kinematically suppressed when the singlet is much lighter than the triplet. Therefore, in order for this annihilation rate to be sufficiently large, $\mdelta=m_{\Sigprime}-m_{\Sprime}$ cannot be too large. 
     The precise value of the upper limit on $\mdelta$ depends on the maximum value of $\lamSigS$ (we take $\lvert \lamSigS \rvert \leq 2$).
  \item Conversely, if  $\lvert\sin \mixAngle \rvert \gtrsim 0.1$, then the DM requirements can also be satisfied by a large $\mdelta$. This occurs because when $\mixAngle$ is large, then the $\Sprime$ can annihilate into weak gauge bosons. This annihilation rate is not sensitive to $\mdelta$. In particular, when $\lvert \sin \mixAngle \rvert \sim 0.1$, the $\Sprime$ can yield the correct relic density without relying on other annihilation channels.
\end{enumerate}

From these observations we conclude that in order for the \xSigmaSM to provide a viable dark matter candidate, the scalar parameters generally fall into one of two categories:
\begin{enumerate}[(a)]
    \item $\lamSigS > 0.1$ and $0 < \mdelta < 30\GeV$.
    \item $\lvert \sin \mixAngle \rvert \gtrsim 0.1 $, $\lamHSigS \gtrsim 0.1$, and $\mdelta>0$.
\end{enumerate}
Aside from the requirements on $\mdelta$, the DM requirements can be satisfied by a wide range of scalar masses. To illustrate this, we select two sets of benchmark scalar couplings and show the mass dependence of the DM relic density and DM constraints in figures~\ref{fig:DM_density}~and~\ref{fig:DM_constraints}, respectively.

\begin{figure}
  \centering
  {
  \includegraphics{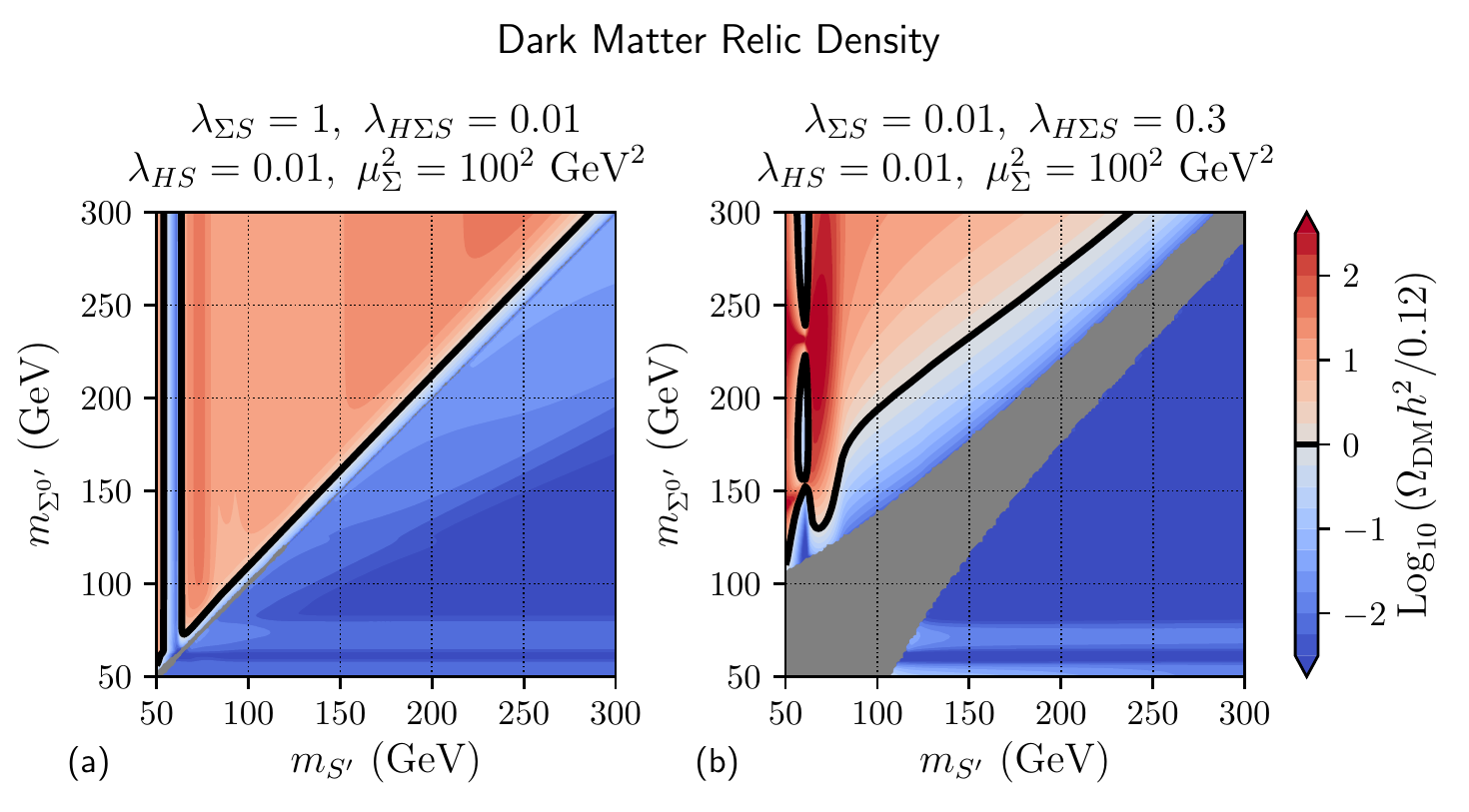}
  \phantomsubcaption\label{fig:DM_density_a}
  \phantomsubcaption\label{fig:DM_density_b}
  }
  \caption{ \label{fig:DM_density}
        The log of the dark matter relic density normalised to the observed, as a function of the masses of the two $m_{\Sprime}$ and $m_{\Sigprime}$ for two sets of benchmark scalar couplings. Both $\lamSig$ and $\lamS$ have negligible impact on dark matter phenomenology and were set to $0.01$. Red regions indicate a DM over-density, while blue regions indicate an under-density. The solid black line is the relic density contour corresponding to the observed relic density. The grey region along $m_{\Sigprime} = m_{\Sprime}$ corresponds to the unphysical region excluded by eq.~\eqref{eq:min_mass_diff}.}
\end{figure}

\begin{figure}
  \centering 
  {
  \includegraphics{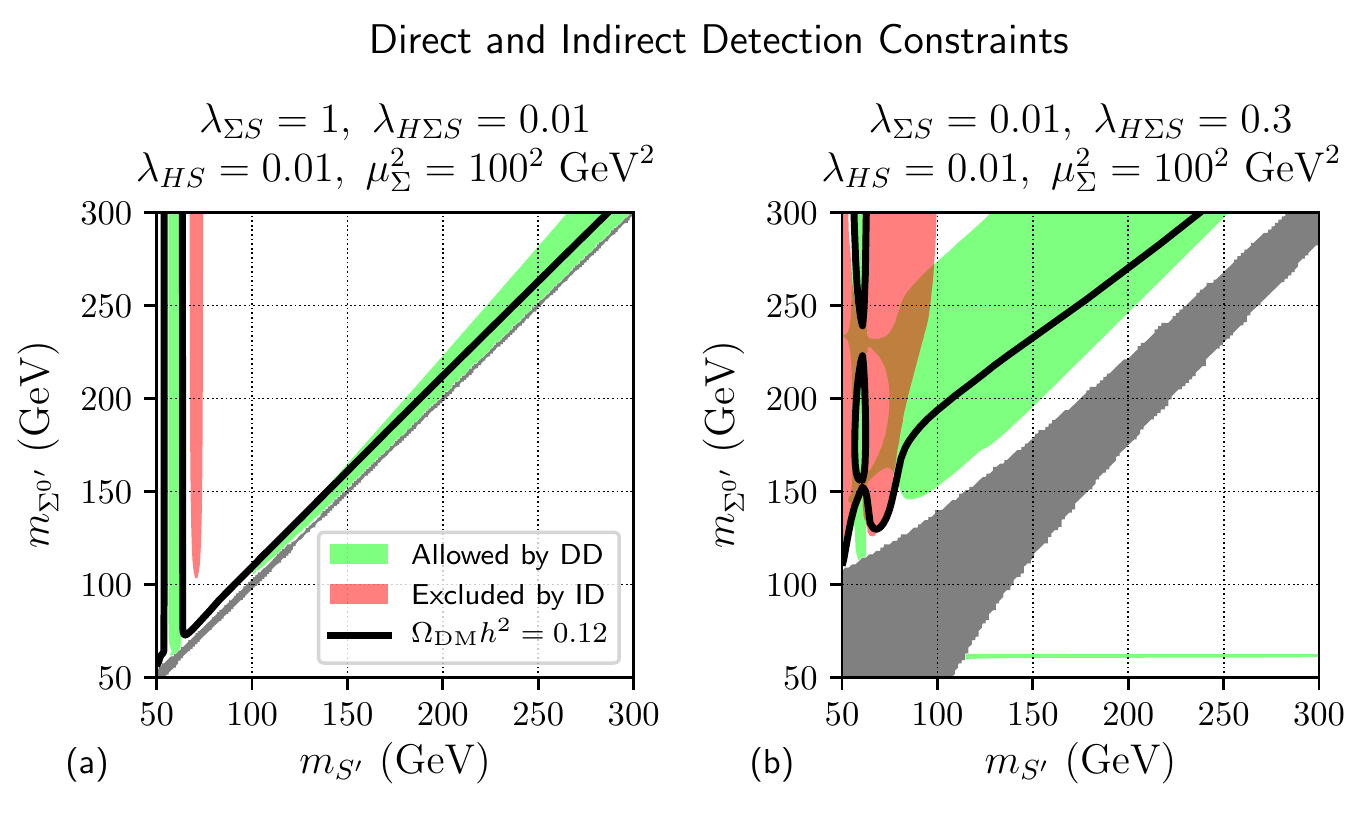}
  \phantomsubcaption\label{fig:DM_constraints_a}
  \phantomsubcaption\label{fig:DM_constraints_b}
  }
  \caption{ \label{fig:DM_constraints}
  The region of parameter space consistent with
  the current direct and indirect detection limits 
  as a function of the masses of the two neutral scalars for two sets of benchmark scalar couplings. Both $\lamSig$ and $\lamS$ have negligible impact on dark matter phenomenology and were set to $0.01$. The
  green region is allowed by direct detection constraints, the red region is excluded by indirect detection, and the black line gives the contour that yields the correct DM density. The grey region along $m_{\Sigprime} = m_{\Sprime}$ corresponds to the unphysical region excluded by eq.~\eqref{eq:min_mass_diff}.
  Electroweak precision observables are not constraining in these regions of parameter space.
  }
\end{figure}

The benchmark couplings used in figures~\ref{fig:DM_density_a}~and~\ref{fig:DM_constraints_a} correspond to the first category, where the correct relic density can only be obtained when $0 <\mdelta \lesssim 30\GeV$, or when $m_{\Sprime} \lesssim \frac{m_H}{2}$. As mentioned previously, this latter region is strongly constrained by invisible Higgs decays and was omitted the from random scan.
The benchmark scalar couplings used in figures~\ref{fig:DM_density_b}~and~\ref{fig:DM_constraints_b} correspond to the second category, where the large neutral scalar large mixing angle allows the correct relic density to be obtained even when $\mdelta$ is large. Once again, the region with $m_{\Sprime} \lesssim \frac{m_H}{2}$ can yield the correct relic density, though in this case the phenomenology of this region is more complex due to the presence of a large scalar mixing angle. In particular, the $\Sprime$--$H$ quartic coupling, which is given by
\begin{equation}
\label{eq:SprimeHiggsQuartic}
\lamHS \cos^2\mixAngle - 
 \lamHSigS \cos \mixAngle \sin \mixAngle + 
 \lamHSig \sin^2 \mixAngle \, ,
\end{equation}
can be equal to zero for some values of $\mixAngle$. While this can eliminate the invisible Higgs decay constraint, it also results in an over-density of DM, such that these cancellations can be clearly seen in figure~\ref{fig:DM_density_b} near $(m_\Sprime, m_\Sigprime) = (60\GeV,150\GeV)$ and $(60\GeV,230\GeV)$.
Furthermore, this region of parameter space is nearly excluded by indirect detection constraints.
Therefore, we will continue to focus on parameter space with $m_\Sprime> \frac{m_H}{2}$.

We have shown that the \xSigmaSM is capable of satisfying DM direct detection constraints while generating the observed DM density for a wide range of scalar couplings and masses. However, as discussed in section~\ref{sec:intro}, pure SU(2) triplet scalar extensions face severe constraints due to collider searches, with a large portion of the triplet masses considered here excluded in minimal triplet extensions. Therefore, in the next section we will examine the collider phenomenology of the $\xSigmaSMmath$.

\section{Collider Phenomenology}
\label{sec:Collider}

The main collider signatures of this model can be summarised as follows:
\begin{itemize}
    \item The new scalars are produced in pairs, primarily through charged and neutral
current Drell-Yan processes. 
    \item They decay via their couplings with other scalars and weak gauge bosons, producing lighter scalars and on- or off-shell SM Higgs and weak gauge bosons. 
    \item Due to the \Ztwo symmetry, the pair production of the new scalar will always result in a final state with at least two stable neutral scalars.\footnote{The $\Sigma^\pm$ is never stable due to the radiative mass splitting of the triplet components.}
    \item Therefore, pair production events will always result in large missing energy alongside decay products from the on- or off-shell SM Higgs and weak gauge bosons.
\end{itemize}
There are several collider searches that are applicable to such signal events. Many of these are supersymmetry searches looking for charginos and neutralinos, with a stable lightest neutralino. This
is a consequence of the fact that the production and decay of $\Sigma^+ \Sigma^-$, $\Sigma^\pm \Sigprime$, and $\Sigma^\pm \Sprime$ pairs is analogous to the production and decay of chargino and neutralino pairs, with large missing energy due to a stable neutralino.

However, there are some notable differences between the scalars and the chargino/neutralinos, such that the constrains obtained in these SUSY searches are not directly applicable:
\begin{itemize}
    \item The production cross section of the charginos/neutralinos is larger than our scalar production cross
section by a factor of $10$--$20$, resulting in significantly weaker constraints.\footnote{This difference is mainly due to kinematic differences between fermions and scalars. For example, the top quark production cross section is similarly larger than that of an equal mass stop.} The leading order production cross sections are shown in
figure~\ref{fig:xsect},
\item The kinematic distributions are slightly different. The scalars are produced with a harder $p_t$ spectrum, which 
generally leads to more stringent bounds.
\item The neutralinos can decay via an (off-shell) $Z$ boson. However, there is no analogous $\Sigprime \rightarrow \Sprime Z^{(*)} $ decay. This is due to the fact that the neutral component of a hypercharge-zero triplet
does not couple to the $Z$.
\end{itemize}
The first point, is not an obstacle to the applicability of chargino/neutralino searches, the cross section limits would still apply and would simply lead to less severe limits on the masses. The effect of the second point is difficult to address without 
using Monte Carlo event generators and re-implementing each of the relevant analyses. The third point is more problematic, as it directly affects the search strategies of the analyses looking for charginos/neutralinos. In order to address the applicability of these chargino/neutralino searches we now discuss the decays of the scalars in more detail.

\begin{figure}
    \centering
    \includegraphics{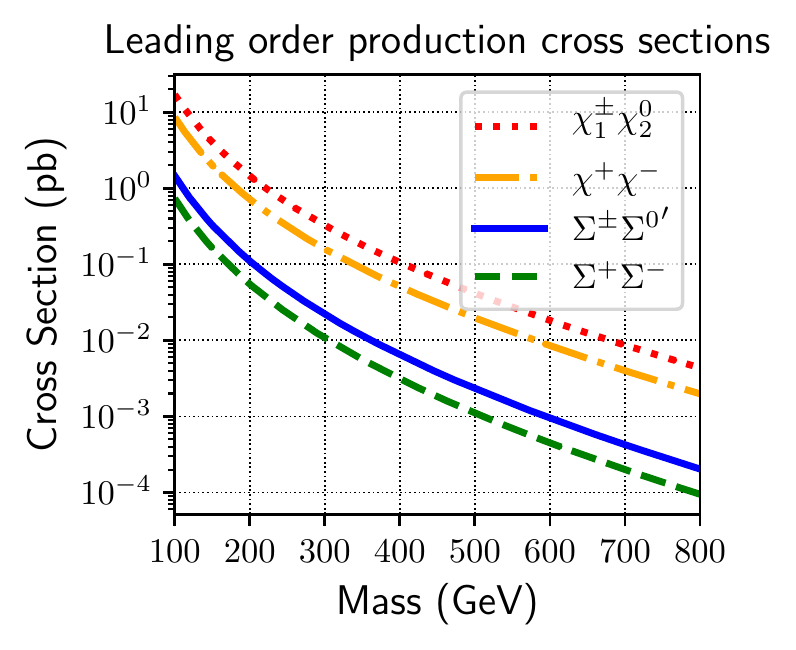}
    \caption{\label{fig:xsect} Leading order production cross sections of $\Sigma^\pm \Sigprime$
      (solid blue line) and $\Sigma^+ \Sigma^-$ (dashed green line) pairs. Note
      that for this plot the scalar mixing angle was set to zero, such that the
      $\Sigma^\pm \Sprime$ production cross section is zero. For comparison, we also show the $\chi^\pm_1 \chi_2^0$ (dotted red line) and $\chi_1^+ \chi_1^-$ (dash-dotted orange
      line) production cross sections. The x-axis corresponds to the mass of the
      pair produced particles, with $m_{\chi_1^\pm} = m_{\chi^0_2}$ and $m_{\Sprime} = m_{\Sigma^\pm}$.
      The scalar cross sections were calculated using \texttt{MadGraph5\_aMC@NLO}~$2.6.5$~f\cite{madgraph}, and the supersymmetric cross sections obtained using $\texttt{Prospino~2.1}$~\cite{prospino_chargino}.}
\end{figure}

\subsection{Decays}
\label{sec:decay}

Motivated by the dark matter phenomenology discussed in section~\ref{sec:dm}, we will focus purely on the parameter space where $\mdelta>0$. Therefore, the $\Sprime$ is always stable, as it is the lightest particle charged under our \Ztwo symmetry. The $\Sigma^+$ can then decay in two ways, via $\Sigma^+ \rightarrow {W^{+}}^{(*)} \Sprime$ or $\Sigma^+  \rightarrow {W^+}^{(*)} \Sigprime$. The former decay is suppressed by the neutral scalar mixing angle, while the latter is only kinematically allowed due to the small radiatively induced mass splitting, $m_{\Sigma^+}-m_\Sigprime \lesssim 166\MeV$. For most of the parameter space, the kinematic suppression is stronger than the mixing angle suppression, such that $\Sigma^\pm$ almost always decays into  ${W^{+}}^{(*)} \Sprime$. As a consequence of there only being one decay channel, the branching ratios of the $\Sigma^+$ decays are not sensitive to the scalar coupling parameters. However, they still determine the $\Sigma^+$ lifetime. 
Fortunately the relevant SUSY searches feature
analogous chargino decays ($\chi_1^+ \rightarrow {W^+}^{(*)} \chi_1^0$). Therefore, except for some differences in the kinematic distributions due to the fermionic nature of charginos, the upper
bound on the production cross section from searches featuring only charginos are directly applicable to $\Sigma^+$.

\begin{figure}
    \centering
    {
    \includegraphics{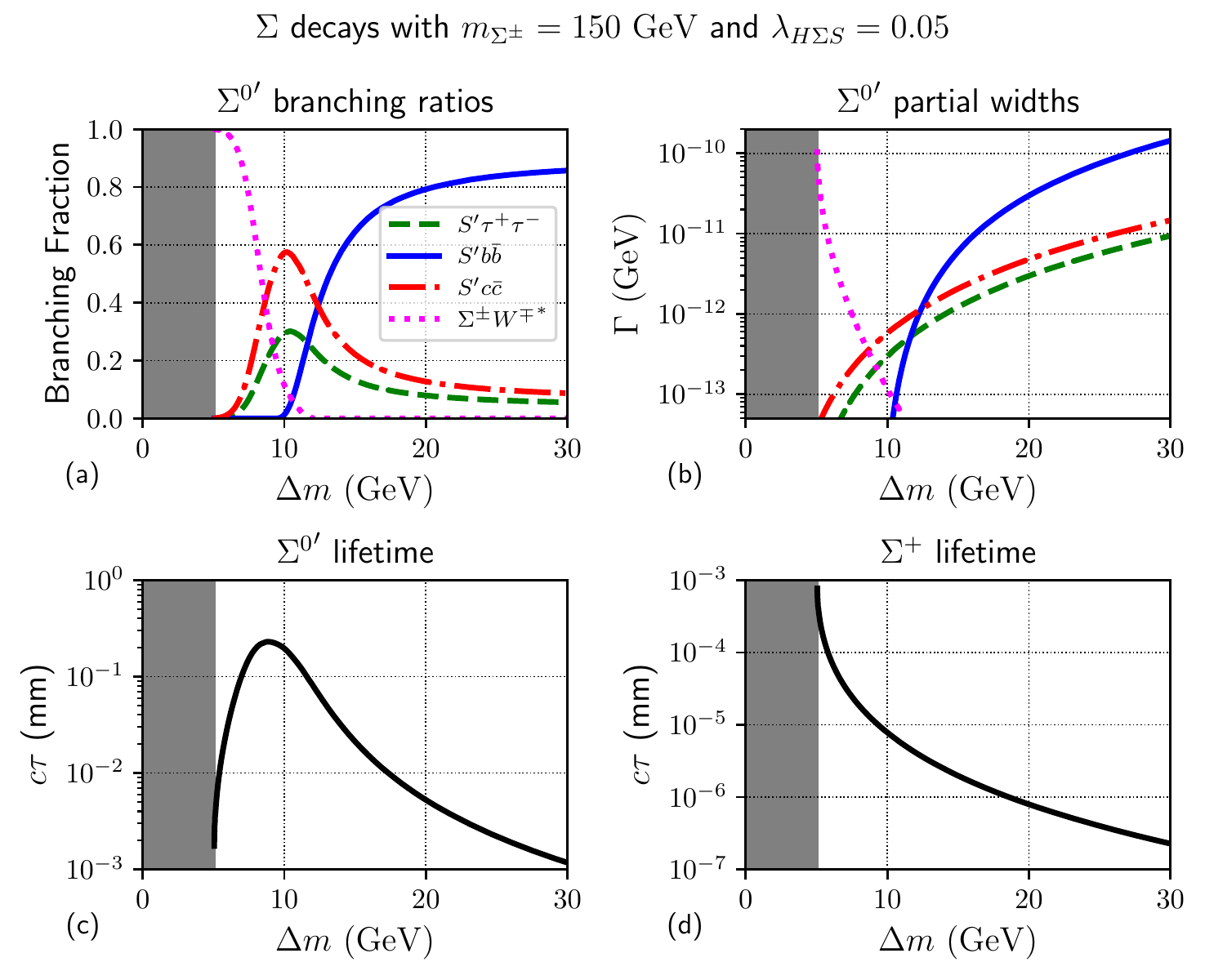}
    \phantomsubcaption\label{fig:decay_a}
    \phantomsubcaption\label{fig:decay_b}
    \phantomsubcaption\label{fig:decay_c}
    \phantomsubcaption\label{fig:decay_d}
    }
    \caption{Plots show the branching ratios (\subref{fig:decay_a}) and partial widths (\subref{fig:decay_b}) of $\Sigprime$ decays, along with the lifetimes of the $\Sigprime$ (\subref{fig:decay_c}) and $\Sigma^+$ (\subref{fig:decay_d}). The branching ratios of the $\Sigma^+$ decays are not shown as they do not change significantly as a function of $\mdelta$. The solid grey area indicates the unphysical region excluded by eq.~\eqref{eq:min_mass_diff}. All scalar couplings were held constant, and $\mdelta$ was varied by changing only $\muS$.}
    \label{fig:decay}
\end{figure}

The decays of the triplet-like
neutral scalar $\Sigprime$ are more complicated. If $\mdelta \gtrsim 10\GeV$, they will predominantly decay into a $\Sprime$ and an (off-shell) SM Higgs. However, for smaller $\mdelta$, they may
instead decay via $\Sigprime \rightarrow \Sigma^\pm {W^{\mp}}^{*}$.
Figure~\ref{fig:decay} shows how the branching ratios and lifetimes of $\Sigprime$ and $\Sigma^+$ vary as a function of $\mdelta$,
for a set of benchmark scalar couplings. The widths were evaluated at leading order using the
\texttt{MadWidth}~\cite{madwidth} component of
\texttt{MadGraph}. Two key features can be seen in figures~\ref{fig:decay_a}~and~\ref{fig:decay_b}:
\begin{itemize}
    \item There is kinematic suppression of the $\Sigprime \rightarrow
\Sprime h^* \rightarrow \Sprime b \bar{b}$ channel for small $\mdelta$. 
    \item The $\Sigprime \rightarrow \Sigma^\pm {W^{\mp}}^{*}$ partial width and branching ratios are large for small $\mdelta$.
\end{itemize}
To explain this latter feature, note that this decay can only take place when
$m_{\Sigprime}>m_{\Sigma^\pm}$. This mass hierarchy is contrary to what is seen in minimal triplet scalar models, where a radiatively induced mass splitting always leads to $m_{\Sigprime} < m_{\Sigma^\pm}$. As is discussed in section~\ref{sec:mass} and illustrated in figure~\ref{fig:mass}, $m_{\Sigprime} > m_{\Sigma^\pm}$ only occurs in our model when the neutral scalar mixing raises the mass of the $m_\Sigprime$ and overcomes this radiative mass splitting. The effect of the scalar mixing is such that $m_{\Sigprime} - m_{\Sigma^\pm}$ is maximised when $\lvert \mdelta \rvert$ is minimised. Therefore, the $\Sigprime \rightarrow \Sigma^\pm {W^\mp}^*$ partial width in figure~\ref{fig:decay_b} is also maximised when $\mdelta$ is minimised. This contrasts with the $\Sigprime \rightarrow \Sprime h^*$ partial widths, which increase with larger $\mdelta$. These competing effects result in the $\Sigprime$ lifetime featuring a local maximum that can be seen in figure~\ref{fig:decay_c}. For this particular set of scalar couplings, the lifetime can be as large as $c\tau \approx 0.2\ \mathrm{mm}$. Given that both of the neutral scalars are stable when $\lamHSigS=0$, the lifetime can be made arbitrarily large by simply decreasing $\lamHSigS$.

Once again comparing the decays to supersymmetric searches, we find that when
the $\mdelta>m_H$, such that $\Sigprime$ decays can produce an
on-shell SM Higgs, there exist collider searches that have analogous decays for
neutralinos. For example, refs.~\cite{cms_chargino_neutralino_higgs_36fb,atlas_chargino_neutralino_higgs_36fb} map
neatly onto our model. Unfortunately the searches relevant to small mass
differences, refs.~\cite{atlas_chargino_neutralino_compressed_140fb,
  cms_chargino_neutralino_compressed_36fb}, have the neutralino decay through an
off-shell $Z$, not an off-shell SM Higgs. While searches using the $Z^*\rightarrow b
\bar{b}$ or $\tau^+ \tau^-$ decay processes could easily be applied to our model, the
relevant decays instead use $Z^* \rightarrow \ell^+ \ell^-$. Therefore
interpreting these results will not be as straightforward.

\subsection{Collider Searches}

Obtaining precise collider limits requires generating events for a
given model, applying the analyses performed in relevant collider searches, and comparing the
predicted number of events with those observed by the relevant searches. Fortunately, there exist useful
tools, such as \texttt{CheckMATE}~\cite{checkmate}, that implement existing
collider searches and make it straightforward to recast their results onto new
models. Unfortunately, the searches most relevant to our model, particularly  refs.~\cite{cms_chargino_neutralino_higgs_36fb,atlas_chargino_neutralino_higgs_36fb,cms_chargino_neutralino_compressed_36fb,cms_chargino_neutralino_diphoton_77fb,atlas_chargino_neutralino_compressed_140fb}, have not yet been
implemented in these tools.\footnote{Some older and superseded searches are
  implemented by these tools, but in a preliminary scan they were not able
  exclude any parameter space with $m_{\Sigprime},m_{\Sigma^+} > 100\GeV$.}
Implementation of all of these analyses requires a considerable amount of work, with some analyses featuring up to $58$ signal regions. For the sake of this initial investigation, we will limit the scope of our collider
phenomenology to re-scaling the bounds on charginos and neutralinos
obtained by these analyses and leave a more thorough investigation as potential future work.

As discussed in the previous section, the dark matter direct detection constraints, together with the requirement that our dark matter candidate constitutes the whole relic density, imply that either the neutral scalar mass difference is small, $0 < \mdelta \lesssim 40\GeV$, or there is a large Higgs coupling $\lamHSigS\gtrsim 0.5$ and mixing angle $\lvert \sin\mixAngle \rvert \gtrsim 0.3$. The two parameter space possibilities are constrained by different collider searches and will be discussed separately with reference to relevant chargino and neutralino searches.

\subsubsection{Small mass difference, $\mdelta < 50\GeV$}
\label{sec:low_diff}

Searching for charginos and neutralinos with compressed mass spectra, mass difference less than $\sim 50\GeV$, is notoriously difficult. The main challenge boils down to the fact that the
visible decay products have low energies due to the limited energy budget, even
when boosted by initial-state radiation jets. Both the CMS~\cite{
  cms_chargino_neutralino_compressed_36fb} and ATLAS~\cite{atlas_chargino_neutralino_compressed_140fb} collaborations have undertaken searches for this type of supersymmetric spectrum, using $36\ifb$ and
$140\ifb$ of data respectively, with the more recent ATLAS search setting more stringent
bounds.

The problem of low-energy visible particles also holds for our scalars. However,
we have the added complication that existing searches use the $\chi_2^0
\rightarrow \chi_1^0 Z^* \rightarrow \chi^0_1 \ell^+ \ell^-$ decay channel. Our
best analogue is the $\Sigprime \rightarrow \Sprime h^* \rightarrow \Sprime
\tau^+ \tau^-$ channel with both of the taus decaying leptonically into a
same-flavor opposite-charge light lepton pair. 
The total branching ratio for this process can be as large as $\sim 2\%$.\footnote{This branching ratio is maximised when the masses and $\lamHSigS$ are selected such that $\mathrm{Br}(\Sigprime \rightarrow \Sigma^\pm {W^\mp}^{*}) = 0$ and 
$\mathrm{Br}(\Sigprime \rightarrow \Sprime \tau^+ \tau^-) \sim 30\%$.} This is smaller than the branching ratio for off-shell $Z$ decays into electrons or muons, which ranges from $7\%$ to $10\%$~\cite{atlas_chargino_neutralino_compressed_140fb}.

Additionally,
relying on the leptonic tau decays to produce light leptons further reduces the
energy of the light leptons, due to some of the $\Sigprime$ decay energy budget
going into the neutrinos produced in the tau decay. Thus, ignoring all other
factors, we would expect weaker limits on the production cross section. This is slightly offset by the fact that the scalar $p_T$ spectrum has a slightly
larger high energy tail than the charginos and neutralinos do. However given
that most of this energy is taken away by the $\Sprime$ it is likely that the
net result is that the model is more difficult to probe.

\begin{figure}
    \centering
    \includegraphics{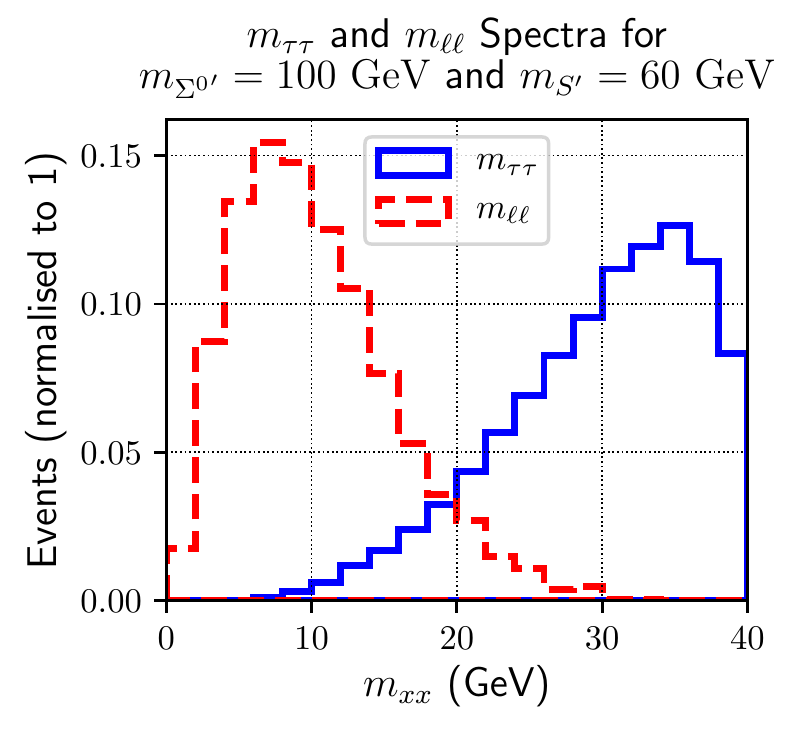}
    \caption{\label{fig:spectrum} Normalised spectrum of tau pair (solid blue line) and light
      lepton pair (dashed red line) invariant masses, where the light leptons
      are the products of leptonic tau decays. This figure is analogous to
      figure~2 in ref.~\cite{atlas_chargino_neutralino_compressed_140fb}, however our
      $m_{\ell \ell}$ spectrum is skewed towards smaller energies.}
\end{figure}

Figure~\ref{fig:spectrum} shows the distribution of the $\tau$ pair and light
lepton-pair invariant masses in the scalar model, obtained using
\texttt{MadGraph}, \texttt{Pythia}~\cite{pythia3}, and
\texttt{MadAnalysis}~\cite{MadAnalysis5}. This figure is analogous to figure~2 in
ref.~\cite{atlas_chargino_neutralino_compressed_140fb}. Note that our
$m_{\tau \tau}$ closely corresponds to the $m_{\ell \ell}$ spectrum computed in
ref.~\cite{atlas_chargino_neutralino_compressed_140fb} for the wino-bino
scenario with $m_{\chi_2^0} \times m_{\chi_1^0} > 0$. However, our actual
$m_{\ell \ell}$ spectrum, where the light leptons arise from leptonic tau
decays, is skewed towards smaller values due to energy taken away by the
neutrinos. The effect for this analysis may be that scalars with a given mass
difference should be compared to charginos and neutralinos with some smaller
mass difference due to the higher proportion of low $m_{\ell \ell}$ events. While this may require us to modify the production cross section limits as a function of $\mdelta$, this should not result in a larger maximal mass reach.

As is discussed in section~\ref{sec:decay}, the $\Sigprime \rightarrow \Sigma^\pm {W^\mp}^{*} \rightarrow \Sprime {W^\pm}^{*} {W^\mp}^{*}$ decay channel branching fraction can be large when $\mdelta \lesssim 10\GeV$. This decay can also produce two opposite-sign same-flavour leptons and contribute signal events. However, as the branching ratio for this decay increases, the $\Sigprime \rightarrow \Sprime \tau^+ \tau^-$ branching ratio decreases. The result is that the total branching ratio to produce opposite-sign same-flavour lepton pairs remains below $7\%$. Furthermore, this channel faces the same challenge of losing some of its limited energy budget to neutrinos. This is compounded by the fact that an appreciable branching fraction necessarily requires $\mdelta$ to be very small. The resulting visible lepton energy spectra are thus shifted towards energies in a manner similar to those arising from leptonic tau decays.

If we ignore the differences in the visible particle $p_T$ spectra, then in
principle one could use the $95\%$ confidence upper bounds on chargino
production reported in ref.~\cite{atlas_chargino_neutralino_compressed_140fb} to
approximate the constraints on our new scalars. Given that our branching fraction to produce the necessary lepton pairs is always smaller than for charginos and neutralinos, we can get conservative constraints by simply
using our scalar production cross sections as shown in
figure~\ref{fig:xsect}.\footnote{Note that whenever the scalar mixing angle is large, the production cross section will be even smaller.} We find that none of the upper limits on the chargino-neutralino production cross section that are provided by the ATLAS analysis~\cite{atlas_chargino_neutralino_compressed_140fb_f41ab,atlas_chargino_neutralino_compressed_140fb_f41cd} are capable of excluding the scalars. Given that we do not expect the visible particle $p_T$ spectrum differences to increase the
mass reach of this analysis, we are confident in saying that the compressed
mass spectrum searches do not currently constrain this model when $m_\Sigprime > 100\GeV$.

\subsubsection{Large mass difference, $\mdelta > m_H$}

When the neutralino mass difference is large, an on-shell SM Higgs boson can be produced in
the decay of neutralinos. Both the CMS~\cite{cms_chargino_neutralino_higgs_36fb}
and ATLAS~\cite{atlas_chargino_neutralino_higgs_36fb} collaborations have
searches looking for this type of decay using $36\ifb$ of data. There are also
some more recent searches looking for diphoton signals in $\chi_2^0\rightarrow
\chi_1^0 h \rightarrow \chi_1^0 \gamma \gamma$, with a $77.5\ifb$ CMS
search~\cite{cms_chargino_neutralino_diphoton_77fb} and a $140\ifb$ ATLAS
search~\cite{atlas_chargino_neutralino_diphoton_140fb}.

Fortunately, unlike the compressed mass spectra searches, these types of
searches are more directly analogous to the decay processes that are present in
our scalar model. Repeating what was done in the previous section and simply
directly using the $\Sigma^\pm \Sigprime$ production cross section shown in
figure~\ref{fig:xsect}, we find that none of the upper-limits provided by these searches
exclude our Type-IV \xSigmaSM. The exclusion boundary lies below the mass ranges considered by ATLAS and CMS. Furthermore, the production cross section in figure~\ref{fig:xsect} is for $\mixAngle=0$, which maximises the number of signal events. Given that DM constraints require $\lvert \sin \mixAngle \gtrsim 0.3$ for $\mdelta \gtrsim 40\GeV$, the actual production cross section for signal events is even smaller.

The more recent $\chi_2^0\rightarrow
\chi_1^0 h \rightarrow \chi_1^0 \gamma \gamma$ searches,
refs.~\cite{cms_chargino_neutralino_diphoton_77fb,
  atlas_chargino_neutralino_diphoton_140fb}, are similarly insensitive. They are
only able to exclude neutralinos with $\chi_1^0 \lesssim 100\GeV$, and have no sensitivity to
scalars with $m_{\Sprime}\sim 100$--$300\GeV$. It should be noted that diphoton
signal regions were included in the $36\ifb$ ATLAS and CMS searches. However,
compared to signal regions with $H\rightarrow b \bar{b} $, they were relatively
insensitive. Upcoming  $H\rightarrow b \bar{b}$ searches with the full Run 2 data set may place stronger constraints on our scalars.

\subsubsection{Intermediate mass difference, $50\GeV < \mdelta < m_H$}

When $\mdelta>50\GeV$, compressed mass spectra
searches become insensitive due to cuts that are designed for low-energy visible
particles. However, the large mass-difference searches with an on-shell SM Higgs
only apply when $\mdelta>m_H$. For weak-scale mass
differences, the chargino and neutralino searches
generally focus on $\chi_2^0 \rightarrow Z^{(*)}
\chi_1^0$~\cite{atlas_chargino_neutralino_Z_140fb}. If the neutralino mass difference is
larger than $m_{Z}$, then these searches generally involve reconstructing an
on-shell $Z$. Therefore, these searches are not applicable to our model due to
the lack of a $\Sigprime \rightarrow Z \Sprime$ decay. If the $Z$ is off-shell, we can use the same approach as in section~\ref{sec:low_diff}. That is, we could once again use the $\Sigprime
\rightarrow \Sprime h^{*}\rightarrow \Sprime \tau \tau$ decay to the generate a reduced number signal events, but the result will be similarly
unconstraining.

One type of analysis we have not yet discussed is a search for chargino pair
production, which looks for $p p \rightarrow \chi_1^+ \chi_1^- \rightarrow
\chi_1^0 \chi_1^0 W^+ W^-$. This is directly analogous to $\Sigma^+ \Sigma^-$
pair production, and therefore such searches could conceivably constrain the scalars when $\mdelta > m_W$. In particular, there is a
recent ATLAS search looking for this type of signal using $140\ifb$ of
data~\cite{atlas_chargino_neutralino_ww_140fb}, which provides upper
limits on the chargino pair production cross
section~\cite{atlas_chargino_neutralino_ww_140fb_t45}. However, once again, interpretation of this data in terms of the $\Sigma^+ \Sigma^-$ pair production cross section results in no limits on the charged triplets.

\subsection{Alternative Searches and Future Prospects}

In addition to the SUSY searches outlined above, there are several other methods that could be employed to search for the $\xSigmaSMmath$. One such method would be to search for displaced vertices. Displaced vertices are the result of a long lived particle that travels a significant distance from the main interaction vertex before decaying. As mentioned in section~\ref{sec:decay}, the lifetime of the $\Sigprime$ can be made arbitrarily large by making $\lamHSigS$ arbitrarily small. Thus, in principle, there are regions of
parameter space that would generate displaced vertices. However, the low
energy of the visible decay products, which is due to the small $\mdelta$, poses a major challenge. Existing displaced vertex searches generally place minimum $p_T$ cuts
that would eliminate any signal events generated by a long lived $\Sigprime$.

Another method of improving the discoverability of the \xSigmaSM would be to eliminate the sub-optimal reliance on chargino and neutralino based searches. In particular, for the compressed mass spectrum scenario, the
fact that existing searches generally look for light lepton pairs coming from an off-shell $Z^*$ decay, means they are ill suited to our model, which relies on $h^*$ decay products. More suitable searches may instead target pairs of low energy $b$ jets, hadronically decaying taus, or taus decaying leptonically into different lepton flavours. The $\Sigprime \rightarrow \Sigma^\pm {W^\mp}^*$ decay, which also has no neutralino analogue, may also allow for alternative searches using $\Sigma^\pm \Sigprime \rightarrow \Sprime \Sprime {W^{\pm}}^{*} {W^{\pm}}^{*} {W^{\mp}}^{*}$ events. However, as was mentioned before, this decay channel is only present when the mass splitting is very small. Also note that a more thorough analysis that relies on this branching ratio at small mass splittings likely requires a more precise treatment of the radiative mass splitting than the approximation we made in section~\ref{sec:mass}. This reliance on chargino and neutralino oriented searches also applies to the intermediate mass difference scenario.

However, even with searches optimised to look for this particular model, the
constraints on these scalars will never be quite as strong as those that
apply to charginos and neutralinos. This is simply due to the fact that the
chargino and neutralino production cross section are roughly a factor of
$10$--$20$ larger than the triplet-scalar production cross section. Therefore, constraints on the scalars with $140\ifb$ of data available are comparable to those that would be obtained in chargino and neutralino searches using $\lesssim 14\ifb$ of data. Given that current chargino and neutralino searches are only just
starting to probe neutralino masses around $100$--$200\GeV$, the
parameter space considered when discussing the dark matter phenomenology in
section~\ref{sec:dm} will likely only begin to be thoroughly probed by analyses using at least $1000\ifb$ of data.

\section{Conclusion}
\label{sec:conclusion}

We have shown that, by introducing a scalar singlet, one is able to relax the collider constraints on minimal SU(2) triplet scalar extensions of the SM. Furthermore, such a model is capable of generating the correct relic density. We have only examined one type of such a model, where both the triplet and singlet
scalars are charged under a single \Ztwo symmetry. There exist several other
variants depending on how \Ztwo charges are assigned. Examination of these
models, or a more robust examination of the collider phenomenology of this model
are potential avenues of further investigation.

This paper focused solely on the collider and dark matter phenomenology, but was motivated by the prospect of novel electroweak phase transitions. The relaxation of collider constraints, combined with the new interactions with the scalar singlet,
opens greater parameter space
for such transitions. 
However, if one requires these new scalars to constitute all of the dark matter density we must impose new constraints, namely that either $m_\Sigprime - m_\Sprime \sim 10$ to $30\GeV$, or $\lamHSigS \gtrsim 0.5 $. 
The natural next step is a detailed examination of the phase transition dynamics to establish if a novel phase transition can indeed be obtained  in some region of this enlarged parameter space.


\appendix
\section{Loop Induced Annihilation into Monochromatic Photons}
\label{app:indirect}
Our dark matter can annihilate into two photons via an off-shell intermediate Higgs boson. This annihilation process has been studied previously in both minimal triplet~\cite{LHCTripDM} and minimal singlet~\cite{Profumo:2010kp,Duerr:2015aka} scalar DM models. The cross section for annihilation into monochromatic photons is given by~\cite{Profumo:2010kp,LHCTripDM,Duerr:2015aka},
\begin{equation}
    \sigma(X X \rightarrow h^* \rightarrow \gamma \gamma) =
    \frac{ \lambda_{H X}^2}{\sqrt{s-4 m_{X}^2}}
    \frac{\Gamma_{h^*\rightarrow \gamma \gamma }(s)}{(s-m_H^2)^2 + m_H^2 \Gamma_h^2} \,,
\end{equation}
where $X$ is either $\Sigprime$ or $\Sprime$, and
\begin{subequations}
\begin{align}
    \lambda_{H \Sprime} &= 
    \lamHS \cos^2(\mixAngle) 
    - 2 \lamHSigS \cos(\mixAngle) \sin(\mixAngle) 
    + \lamHSig \sin^2(\mixAngle) \, ,\\
    \lambda_{H \Sigprime} &= 
    \lamHSig \cos^2(\mixAngle) 
    + 2 \lamHSigS \cos(\mixAngle) \sin(\mixAngle) 
    + \lamHS \sin^2(\mixAngle) \, .
\end{align}
\end{subequations}
The off-shell Higgs diphoton rate is given by,
\begin{equation}
\begin{aligned}
\Gamma_{h^*\rightarrow \gamma \gamma}(s) = \frac{\alpha^2 s^{3/2}}{256 \pi^3 v_H^2} & \left\lvert \, \frac{4}{3} A_{1/2}\left(\frac{s}{4 m_t}\right) 
+  A_{1}\left(\frac{s}{4 m_w}\right) \right. \\
&\left.  + \frac{2 v_H \lamHSig m_W }{g m_{\Sigma^+}^2} A_{0}
\left(\frac{s}{4 m_{\Sigma^+}}\right)  \right\rvert ^2 \, ,
\end{aligned}
\end{equation}
with the loop functions, 
\begin{subequations}
\begin{align}
    A_0(x) &= - (x - f(x))x^{-2} \, ,\\
    A_{1/2}(x) &= 2 (x + (x-1)f(x))x^{-2} \, ,\\
    A_{1}(x) &= -(3x + 2x^2 +3(2 x -1)f(x))x^{-2} \, ,\\
    f(x) & =
           \begin{cases} 
             \mrm{arcsin}^2 (\sqrt{x}) & x\leq 1 \\
             - \frac{1}{4}\left( \ln\frac{1 + \sqrt{1-x^{-1}}}{1-\sqrt{1-x^{-1}}} - i \pi \right)^2 &  x > 1 \\
           \end{cases}
    \, .
\end{align}
\end{subequations}
If we set $\mixAngle=0$ and make the triplet very heavy, our diphoton annihilation rates are consistent with the minimal singlet scalar DM annihilation rates in refs.~\cite{Profumo:2010kp}~and~\cite{Duerr:2015aka}.


In order to get our Fermi-LAT constraint we then evaluate $\sigma v$ in the zero-velocity limit ($s\rightarrow 4 m_{X}^2$). If our model results in a DM under-density, such that it only contributes a fraction of the DM density, we scale the annihilation rate by $\left(\frac{\Omega_{\mrm{DM}}h^2}{0.12}\right)^2$.

\bibliography{main}

\providecommand{\href}[2]{#2}\begingroup\raggedright\begin{thebibliography}{100}

\bibitem{TroddenReview}
M.~Trodden, \emph{{Electroweak baryogenesis: A Brief review}},  in
  \emph{{Proceedings, 33rd Rencontres de Moriond 98 electrowek interactions and
  unified theories: Les Arcs, France, Mar 14-21, 1998}}, pp.~471--480, 1998,
  \href{https://arxiv.org/abs/hep-ph/9805252}{{\ttfamily hep-ph/9805252}}.

\bibitem{MorrisseyMJRMReview}
D.~E. Morrissey and M.~J. Ramsey-Musolf, \emph{{Electroweak baryogenesis}},
  \href{https://doi.org/10.1088/1367-2630/14/12/125003}{\emph{New J. Phys.}
  {\bfseries 14} (2012) 125003}
  [\href{https://arxiv.org/abs/1206.2942}{{\ttfamily 1206.2942}}].

\bibitem{Weir:2017wfa}
D.~J. Weir, \emph{{Gravitational waves from a first order electroweak phase
  transition: a brief review}},
  \href{https://doi.org/10.1098/rsta.2017.0126}{\emph{Phil. Trans. Roy. Soc.
  Lond. A} {\bfseries 376} (2018) 20170126}
  [\href{https://arxiv.org/abs/1705.01783}{{\ttfamily 1705.01783}}].

\bibitem{GrahamReview}
A.~Mazumdar and G.~White, \emph{{Cosmic phase transitions: their applications
  and experimental signatures}},
  \href{https://arxiv.org/abs/1811.01948}{{\ttfamily 1811.01948}}.

\bibitem{Ramsey-Musolf:2019lsf}
M.~J. Ramsey-Musolf, \emph{{The Electroweak Phase Transition: A Collider
  Target}},  \href{https://arxiv.org/abs/1912.07189}{{\ttfamily 1912.07189}}.

\bibitem{Kajantie:1996qd}
K.~Kajantie, M.~Laine, K.~Rummukainen and M.~E. Shaposhnikov, \emph{{A
  Nonperturbative analysis of the finite T phase transition in SU(2) x U(1)
  electroweak theory}},
  \href{https://doi.org/10.1016/S0550-3213(97)00164-8}{\emph{Nucl. Phys.}
  {\bfseries B493} (1997) 413}
  [\href{https://arxiv.org/abs/hep-lat/9612006}{{\ttfamily hep-lat/9612006}}].

\bibitem{Kajantie:1996mn}
K.~Kajantie, M.~Laine, K.~Rummukainen and M.~E. Shaposhnikov, \emph{{Is there a
  hot electroweak phase transition at m(H) larger or equal to m(W)?}},
  \href{https://doi.org/10.1103/PhysRevLett.77.2887}{\emph{Phys. Rev. Lett.}
  {\bfseries 77} (1996) 2887}
  [\href{https://arxiv.org/abs/hep-ph/9605288}{{\ttfamily hep-ph/9605288}}].

\bibitem{Gurtler:1997hr}
M.~Gurtler, E.-M. Ilgenfritz and A.~Schiller, \emph{{Where the electroweak
  phase transition ends}},
  \href{https://doi.org/10.1103/PhysRevD.56.3888}{\emph{Phys. Rev.} {\bfseries
  D56} (1997) 3888} [\href{https://arxiv.org/abs/hep-lat/9704013}{{\ttfamily
  hep-lat/9704013}}].

\bibitem{Laine:1998jb}
M.~Laine and K.~Rummukainen, \emph{{What's new with the electroweak phase
  transition?}},
  \href{https://doi.org/10.1016/S0920-5632(99)85017-8}{\emph{Nucl. Phys. Proc.
  Suppl.} {\bfseries 73} (1999) 180}
  [\href{https://arxiv.org/abs/hep-lat/9809045}{{\ttfamily hep-lat/9809045}}].

\bibitem{Csikor:1998eu}
F.~Csikor, Z.~Fodor and J.~Heitger, \emph{{Endpoint of the hot electroweak
  phase transition}},
  \href{https://doi.org/10.1103/PhysRevLett.82.21}{\emph{Phys. Rev. Lett.}
  {\bfseries 82} (1999) 21}
  [\href{https://arxiv.org/abs/hep-ph/9809291}{{\ttfamily hep-ph/9809291}}].

\bibitem{Aoki:1999fi}
Y.~Aoki, F.~Csikor, Z.~Fodor and A.~Ukawa, \emph{{The Endpoint of the first
  order phase transition of the SU(2) gauge Higgs model on a four-dimensional
  isotropic lattice}},
  \href{https://doi.org/10.1103/PhysRevD.60.013001}{\emph{Phys. Rev.}
  {\bfseries D60} (1999) 013001}
  [\href{https://arxiv.org/abs/hep-lat/9901021}{{\ttfamily hep-lat/9901021}}].

\bibitem{Espinosa:1993bs}
J.~Espinosa and M.~Quiros, \emph{{The Electroweak phase transition with a
  singlet}}, \href{https://doi.org/10.1016/0370-2693(93)91111-Y}{\emph{Phys.
  Lett. B} {\bfseries 305} (1993) 98}
  [\href{https://arxiv.org/abs/hep-ph/9301285}{{\ttfamily hep-ph/9301285}}].

\bibitem{Benson:1993qx}
K.~E. Benson, \emph{{Avoiding baryon washout in the extended Standard Model}},
  \href{https://doi.org/10.1103/PhysRevD.48.2456}{\emph{Phys. Rev. D}
  {\bfseries 48} (1993) 2456}.

\bibitem{Choi:1993cv}
J.~Choi and R.~Volkas, \emph{{Real Higgs singlet and the electroweak phase
  transition in the Standard Model}},
  \href{https://doi.org/10.1016/0370-2693(93)91013-D}{\emph{Phys. Lett. B}
  {\bfseries 317} (1993) 385}
  [\href{https://arxiv.org/abs/hep-ph/9308234}{{\ttfamily hep-ph/9308234}}].

\bibitem{Vergara:1996ub}
L.~Vergara, \emph{{Baryon asymmetry persistence in the standard model with a
  singlet}}, \href{https://doi.org/10.1103/PhysRevD.55.5248}{\emph{Phys. Rev.
  D} {\bfseries 55} (1997) 5248}.

\bibitem{Ham:2004cf}
S.~Ham, Y.~Jeong and S.~Oh, \emph{{Electroweak phase transition in an extension
  of the standard model with a real Higgs singlet}},
  \href{https://doi.org/10.1088/0954-3899/31/8/017}{\emph{J. Phys. G}
  {\bfseries 31} (2005) 857}
  [\href{https://arxiv.org/abs/hep-ph/0411352}{{\ttfamily hep-ph/0411352}}].

\bibitem{Ahriche:2007jp}
A.~Ahriche, \emph{{What is the criterion for a strong first order electroweak
  phase transition in singlet models?}},
  \href{https://doi.org/10.1103/PhysRevD.75.083522}{\emph{Phys. Rev.}
  {\bfseries D75} (2007) 083522}
  [\href{https://arxiv.org/abs/hep-ph/0701192}{{\ttfamily hep-ph/0701192}}].

\bibitem{Profumo:2007wc}
S.~Profumo, M.~J. Ramsey-Musolf and G.~Shaughnessy, \emph{{Singlet Higgs
  phenomenology and the electroweak phase transition}},
  \href{https://doi.org/10.1088/1126-6708/2007/08/010}{\emph{JHEP} {\bfseries
  08} (2007) 010} [\href{https://arxiv.org/abs/0705.2425}{{\ttfamily
  0705.2425}}].

\bibitem{Noble:2007kk}
A.~Noble and M.~Perelstein, \emph{{Higgs self-coupling as a probe of
  electroweak phase transition}},
  \href{https://doi.org/10.1103/PhysRevD.78.063518}{\emph{Phys. Rev.}
  {\bfseries D78} (2008) 063518}
  [\href{https://arxiv.org/abs/0711.3018}{{\ttfamily 0711.3018}}].

\bibitem{Espinosa:2007qk}
J.~R. Espinosa and M.~Quiros, \emph{{Novel Effects in Electroweak Breaking from
  a Hidden Sector}},
  \href{https://doi.org/10.1103/PhysRevD.76.076004}{\emph{Phys. Rev. D}
  {\bfseries 76} (2007) 076004}
  [\href{https://arxiv.org/abs/hep-ph/0701145}{{\ttfamily hep-ph/0701145}}].

\bibitem{Espinosa:2008kw}
J.~Espinosa, T.~Konstandin, J.~No and M.~Quiros, \emph{{Some Cosmological
  Implications of Hidden Sectors}},
  \href{https://doi.org/10.1103/PhysRevD.78.123528}{\emph{Phys. Rev. D}
  {\bfseries 78} (2008) 123528}
  [\href{https://arxiv.org/abs/0809.3215}{{\ttfamily 0809.3215}}].

\bibitem{Barger:2007im}
V.~Barger, P.~Langacker, M.~McCaskey, M.~J. Ramsey-Musolf and G.~Shaughnessy,
  \emph{{LHC Phenomenology of an Extended Standard Model with a Real Scalar
  Singlet}}, \href{https://doi.org/10.1103/PhysRevD.77.035005}{\emph{Phys.
  Rev.} {\bfseries D77} (2008) 035005}
  [\href{https://arxiv.org/abs/0706.4311}{{\ttfamily 0706.4311}}].

\bibitem{Ashoorioon:2009nf}
A.~Ashoorioon and T.~Konstandin, \emph{{Strong electroweak phase transitions
  without collider traces}},
  \href{https://doi.org/10.1088/1126-6708/2009/07/086}{\emph{JHEP} {\bfseries
  07} (2009) 086} [\href{https://arxiv.org/abs/0904.0353}{{\ttfamily
  0904.0353}}].

\bibitem{Das:2009ue}
S.~Das, P.~J. Fox, A.~Kumar and N.~Weiner, \emph{{The Dark Side of the
  Electroweak Phase Transition}},
  \href{https://doi.org/10.1007/JHEP11(2010)108}{\emph{JHEP} {\bfseries 11}
  (2010) 108} [\href{https://arxiv.org/abs/0910.1262}{{\ttfamily 0910.1262}}].

\bibitem{Espinosa:2011ax}
J.~R. Espinosa, T.~Konstandin and F.~Riva, \emph{{Strong Electroweak Phase
  Transitions in the Standard Model with a Singlet}},
  \href{https://doi.org/10.1016/j.nuclphysb.2011.09.010}{\emph{Nucl. Phys.}
  {\bfseries B854} (2012) 592}
  [\href{https://arxiv.org/abs/1107.5441}{{\ttfamily 1107.5441}}].

\bibitem{Cline:2012hg}
J.~M. Cline and K.~Kainulainen, \emph{{Electroweak baryogenesis and dark matter
  from a singlet Higgs}},
  \href{https://doi.org/10.1088/1475-7516/2013/01/012}{\emph{JCAP} {\bfseries
  1301} (2013) 012} [\href{https://arxiv.org/abs/1210.4196}{{\ttfamily
  1210.4196}}].

\bibitem{Chung:2012vg}
D.~J.~H. Chung, A.~J. Long and L.-T. Wang, \emph{{125 GeV Higgs boson and
  electroweak phase transition model classes}},
  \href{https://doi.org/10.1103/PhysRevD.87.023509}{\emph{Phys. Rev.}
  {\bfseries D87} (2013) 023509}
  [\href{https://arxiv.org/abs/1209.1819}{{\ttfamily 1209.1819}}].

\bibitem{Barger:2011vm}
V.~Barger, D.~J. Chung, A.~J. Long and L.-T. Wang, \emph{{Strongly First Order
  Phase Transitions Near an Enhanced Discrete Symmetry Point}},
  \href{https://doi.org/10.1016/j.physletb.2012.02.040}{\emph{Phys. Lett. B}
  {\bfseries 710} (2012) 1} [\href{https://arxiv.org/abs/1112.5460}{{\ttfamily
  1112.5460}}].

\bibitem{Huang:2012wn}
W.~Huang, J.~Shu and Y.~Zhang, \emph{{On the Higgs Fit and Electroweak Phase
  Transition}}, \href{https://doi.org/10.1007/JHEP03(2013)164}{\emph{JHEP}
  {\bfseries 03} (2013) 164} [\href{https://arxiv.org/abs/1210.0906}{{\ttfamily
  1210.0906}}].

\bibitem{Damgaard:2013kva}
P.~H. Damgaard, D.~O'Connell, T.~C. Petersen and A.~Tranberg,
  \emph{{Constraints on New Physics from Baryogenesis and Large Hadron Collider
  Data}}, \href{https://doi.org/10.1103/PhysRevLett.111.221804}{\emph{Phys.
  Rev. Lett.} {\bfseries 111} (2013) 221804}
  [\href{https://arxiv.org/abs/1305.4362}{{\ttfamily 1305.4362}}].

\bibitem{Fairbairn:2013uta}
M.~Fairbairn and R.~Hogan, \emph{{Singlet Fermionic Dark Matter and the
  Electroweak Phase Transition}},
  \href{https://doi.org/10.1007/JHEP09(2013)022}{\emph{JHEP} {\bfseries 09}
  (2013) 022} [\href{https://arxiv.org/abs/1305.3452}{{\ttfamily 1305.3452}}].

\bibitem{No:2013wsa}
J.~M. No and M.~Ramsey-Musolf, \emph{{Probing the Higgs Portal at the LHC
  Through Resonant di-Higgs Production}},
  \href{https://doi.org/10.1103/PhysRevD.89.095031}{\emph{Phys. Rev. D}
  {\bfseries 89} (2014) 095031}
  [\href{https://arxiv.org/abs/1310.6035}{{\ttfamily 1310.6035}}].

\bibitem{Profumo:2014opa}
S.~Profumo, M.~J. Ramsey-Musolf, C.~L. Wainwright and P.~Winslow,
  \emph{{Singlet-catalyzed electroweak phase transitions and precision Higgs
  boson studies}},
  \href{https://doi.org/10.1103/PhysRevD.91.035018}{\emph{Phys. Rev.}
  {\bfseries D91} (2015) 035018}
  [\href{https://arxiv.org/abs/1407.5342}{{\ttfamily 1407.5342}}].

\bibitem{Craig:2014lda}
N.~Craig, H.~K. Lou, M.~McCullough and A.~Thalapillil, \emph{{The Higgs Portal
  Above Threshold}}, \href{https://doi.org/10.1007/JHEP02(2016)127}{\emph{JHEP}
  {\bfseries 02} (2016) 127} [\href{https://arxiv.org/abs/1412.0258}{{\ttfamily
  1412.0258}}].

\bibitem{Curtin:2014jma}
D.~Curtin, P.~Meade and C.-T. Yu, \emph{{Testing Electroweak Baryogenesis with
  Future Colliders}},
  \href{https://doi.org/10.1007/JHEP11(2014)127}{\emph{JHEP} {\bfseries 11}
  (2014) 127} [\href{https://arxiv.org/abs/1409.0005}{{\ttfamily 1409.0005}}].

\bibitem{Chen:2014ask}
C.-Y. Chen, S.~Dawson and I.~M. Lewis, \emph{{Exploring resonant di-Higgs boson
  production in the Higgs singlet model}},
  \href{https://doi.org/10.1103/PhysRevD.91.035015}{\emph{Phys. Rev.}
  {\bfseries D91} (2015) 035015}
  [\href{https://arxiv.org/abs/1410.5488}{{\ttfamily 1410.5488}}].

\bibitem{Katz:2014bha}
A.~Katz and M.~Perelstein, \emph{{Higgs Couplings and Electroweak Phase
  Transition}}, \href{https://doi.org/10.1007/JHEP07(2014)108}{\emph{JHEP}
  {\bfseries 07} (2014) 108} [\href{https://arxiv.org/abs/1401.1827}{{\ttfamily
  1401.1827}}].

\bibitem{Kozaczuk:2015owa}
J.~Kozaczuk, \emph{{Bubble Expansion and the Viability of Singlet-Driven
  Electroweak Baryogenesis}},
  \href{https://doi.org/10.1007/JHEP10(2015)135}{\emph{JHEP} {\bfseries 10}
  (2015) 135} [\href{https://arxiv.org/abs/1506.04741}{{\ttfamily
  1506.04741}}].

\bibitem{Kanemura:2015fra}
S.~Kanemura, M.~Kikuchi and K.~Yagyu, \emph{{Radiative corrections to the Higgs
  boson couplings in the model with an additional real singlet scalar field}},
  \href{https://doi.org/10.1016/j.nuclphysb.2016.04.005}{\emph{Nucl. Phys.}
  {\bfseries B907} (2016) 286}
  [\href{https://arxiv.org/abs/1511.06211}{{\ttfamily 1511.06211}}].

\bibitem{Damgaard:2015con}
P.~H. Damgaard, A.~Haarr, D.~O'Connell and A.~Tranberg, \emph{{Effective Field
  Theory and Electroweak Baryogenesis in the Singlet-Extended Standard Model}},
  \href{https://doi.org/10.1007/JHEP02(2016)107}{\emph{JHEP} {\bfseries 02}
  (2016) 107} [\href{https://arxiv.org/abs/1512.01963}{{\ttfamily
  1512.01963}}].

\bibitem{Huang:2015tdv}
P.~Huang, A.~Joglekar, B.~Li and C.~E.~M. Wagner, \emph{{Probing the
  Electroweak Phase Transition at the LHC}},
  \href{https://doi.org/10.1103/PhysRevD.93.055049}{\emph{Phys. Rev.}
  {\bfseries D93} (2016) 055049}
  [\href{https://arxiv.org/abs/1512.00068}{{\ttfamily 1512.00068}}].

\bibitem{Kanemura:2016lkz}
S.~Kanemura, M.~Kikuchi and K.~Yagyu, \emph{{One-loop corrections to the Higgs
  self-couplings in the singlet extension}},
  \href{https://doi.org/10.1016/j.nuclphysb.2017.02.004}{\emph{Nucl. Phys.}
  {\bfseries B917} (2017) 154}
  [\href{https://arxiv.org/abs/1608.01582}{{\ttfamily 1608.01582}}].

\bibitem{Kotwal:2016tex}
A.~V. Kotwal, M.~J. Ramsey-Musolf, J.~M. No and P.~Winslow,
  \emph{{Singlet-catalyzed electroweak phase transitions in the 100 TeV
  frontier}}, \href{https://doi.org/10.1103/PhysRevD.94.035022}{\emph{Phys.
  Rev.} {\bfseries D94} (2016) 035022}
  [\href{https://arxiv.org/abs/1605.06123}{{\ttfamily 1605.06123}}].

\bibitem{Brauner:2016fla}
T.~Brauner, T.~V.~I. Tenkanen, A.~Tranberg, A.~Vuorinen and D.~J. Weir,
  \emph{{Dimensional reduction of the Standard Model coupled to a new singlet
  scalar field}}, \href{https://doi.org/10.1007/JHEP03(2017)007}{\emph{JHEP}
  {\bfseries 03} (2017) 007}
  [\href{https://arxiv.org/abs/1609.06230}{{\ttfamily 1609.06230}}].

\bibitem{Huang:2017jws}
T.~Huang, J.~No, L.~Pernié, M.~Ramsey-Musolf, A.~Safonov, M.~Spannowsky
  et~al., \emph{{Resonant di-Higgs boson production in the $b{\bar{b}}WW$
  channel: Probing the electroweak phase transition at the LHC}},
  \href{https://doi.org/10.1103/PhysRevD.96.035007}{\emph{Phys. Rev. D}
  {\bfseries 96} (2017) 035007}
  [\href{https://arxiv.org/abs/1701.04442}{{\ttfamily 1701.04442}}].

\bibitem{Chen:2017qcz}
C.-Y. Chen, J.~Kozaczuk and I.~M. Lewis, \emph{{Non-resonant Collider
  Signatures of a Singlet-Driven Electroweak Phase Transition}},
  \href{https://doi.org/10.1007/JHEP08(2017)096}{\emph{JHEP} {\bfseries 08}
  (2017) 096} [\href{https://arxiv.org/abs/1704.05844}{{\ttfamily
  1704.05844}}].

\bibitem{Beniwal:2017eik}
A.~Beniwal, M.~Lewicki, J.~D. Wells, M.~White and A.~G. Williams,
  \emph{{Gravitational wave, collider and dark matter signals from a scalar
  singlet electroweak baryogenesis}},
  \href{https://doi.org/10.1007/JHEP08(2017)108}{\emph{JHEP} {\bfseries 08}
  (2017) 108} [\href{https://arxiv.org/abs/1702.06124}{{\ttfamily
  1702.06124}}].

\bibitem{Cline:2017qpe}
J.~M. Cline, K.~Kainulainen and D.~Tucker-Smith, \emph{{Electroweak
  baryogenesis from a dark sector}},
  \href{https://doi.org/10.1103/PhysRevD.95.115006}{\emph{Phys. Rev. D}
  {\bfseries 95} (2017) 115006}
  [\href{https://arxiv.org/abs/1702.08909}{{\ttfamily 1702.08909}}].

\bibitem{Kurup:2017dzf}
G.~Kurup and M.~Perelstein, \emph{{Dynamics of Electroweak Phase Transition In
  Singlet-Scalar Extension of the Standard Model}},
  \href{https://doi.org/10.1103/PhysRevD.96.015036}{\emph{Phys. Rev.}
  {\bfseries D96} (2017) 015036}
  [\href{https://arxiv.org/abs/1704.03381}{{\ttfamily 1704.03381}}].

\bibitem{Alves:2018jsw}
A.~Alves, T.~Ghosh, H.-K. Guo, K.~Sinha and D.~Vagie, \emph{{Collider and
  Gravitational Wave Complementarity in Exploring the Singlet Extension of the
  Standard Model}},  \href{https://arxiv.org/abs/1812.09333}{{\ttfamily
  1812.09333}}.

\bibitem{Li:2019tfd}
H.-L. Li, M.~Ramsey-Musolf and S.~Willocq, \emph{{Probing a scalar
  singlet-catalyzed electroweak phase transition with resonant di-Higgs boson
  production in the $4b$ channel}},
  \href{https://doi.org/10.1103/PhysRevD.100.075035}{\emph{Phys. Rev.}
  {\bfseries D100} (2019) 075035}
  [\href{https://arxiv.org/abs/1906.05289}{{\ttfamily 1906.05289}}].

\bibitem{Gould:2019qek}
O.~Gould, J.~Kozaczuk, L.~Niemi, M.~J. Ramsey-Musolf, T.~V.~I. Tenkanen and
  D.~J. Weir, \emph{{Nonperturbative analysis of the gravitational waves from a
  first-order electroweak phase transition}},
  \href{https://arxiv.org/abs/1903.11604}{{\ttfamily 1903.11604}}.

\bibitem{Kozaczuk:2019pet}
J.~Kozaczuk, M.~J. Ramsey-Musolf and J.~Shelton, \emph{{Exotic Higgs Decays and
  the Electroweak Phase Transition}},
  \href{https://arxiv.org/abs/1911.10210}{{\ttfamily 1911.10210}}.

\bibitem{Carena:2019une}
M.~Carena, Z.~Liu and Y.~Wang, \emph{{Electroweak $\!$phase $\!$transition
  $\!$with $\!$spontaneous $\!$$Z_2$-breaking}},
  \href{https://arxiv.org/abs/1911.10206}{{\ttfamily 1911.10206}}.

\bibitem{Heinemann:2019trx}
B.~Heinemann and Y.~Nir, \emph{{The Higgs program and open questions in
  particle physics and cosmology}},
  \href{https://doi.org/10.3367/UFNe.2019.05.038568}{\emph{Usp. Fiz. Nauk}
  {\bfseries 189} (2019) 985}
  [\href{https://arxiv.org/abs/1905.00382}{{\ttfamily 1905.00382}}].

\bibitem{Turok:1991uc}
N.~Turok and J.~Zadrozny, \emph{{Phase transitions in the two doublet model}},
  \href{https://doi.org/10.1016/0550-3213(92)90284-I}{\emph{Nucl. Phys. B}
  {\bfseries 369} (1992) 729}.

\bibitem{Davies:1994id}
A.~Davies, C.~Froggatt, G.~Jenkins and R.~Moorhouse, \emph{{Baryogenesis
  constraints on two Higgs doublet models}},
  \href{https://doi.org/10.1016/0370-2693(94)90559-2}{\emph{Phys. Lett. B}
  {\bfseries 336} (1994) 464}.

\bibitem{Hammerschmitt:1994fn}
A.~Hammerschmitt, J.~Kripfganz and M.~G. Schmidt, \emph{{Baryon asymmetry from
  a two stage electroweak phase transition?}},
  \href{https://doi.org/10.1007/BF01557241}{\emph{Z. Phys.} {\bfseries C64}
  (1994) 105} [\href{https://arxiv.org/abs/hep-ph/9404272}{{\ttfamily
  hep-ph/9404272}}].

\bibitem{Cline:1996mga}
J.~M. Cline and P.-A. Lemieux, \emph{{Electroweak phase transition in two Higgs
  doublet models}}, \href{https://doi.org/10.1103/PhysRevD.55.3873}{\emph{Phys.
  Rev. D} {\bfseries 55} (1997) 3873}
  [\href{https://arxiv.org/abs/hep-ph/9609240}{{\ttfamily hep-ph/9609240}}].

\bibitem{Fromme:2006cm}
L.~Fromme, S.~J. Huber and M.~Seniuch, \emph{{Baryogenesis in the two-Higgs
  doublet model}},
  \href{https://doi.org/10.1088/1126-6708/2006/11/038}{\emph{JHEP} {\bfseries
  11} (2006) 038} [\href{https://arxiv.org/abs/hep-ph/0605242}{{\ttfamily
  hep-ph/0605242}}].

\bibitem{Cline:2011mm}
J.~M. Cline, K.~Kainulainen and M.~Trott, \emph{{Electroweak Baryogenesis in
  Two Higgs Doublet Models and B meson anomalies}},
  \href{https://doi.org/10.1007/JHEP11(2011)089}{\emph{JHEP} {\bfseries 11}
  (2011) 089} [\href{https://arxiv.org/abs/1107.3559}{{\ttfamily 1107.3559}}].

\bibitem{Dorsch:2013wja}
G.~Dorsch, S.~Huber and J.~No, \emph{{A strong electroweak phase transition in
  the 2HDM after LHC8}},
  \href{https://doi.org/10.1007/JHEP10(2013)029}{\emph{JHEP} {\bfseries 10}
  (2013) 029} [\href{https://arxiv.org/abs/1305.6610}{{\ttfamily 1305.6610}}].

\bibitem{Dorsch:2014qja}
G.~Dorsch, S.~Huber, K.~Mimasu and J.~No, \emph{{Echoes of the Electroweak
  Phase Transition: Discovering a second Higgs doublet through $A_0 \rightarrow
  ZH_0$}}, \href{https://doi.org/10.1103/PhysRevLett.113.211802}{\emph{Phys.
  Rev. Lett.} {\bfseries 113} (2014) 211802}
  [\href{https://arxiv.org/abs/1405.5537}{{\ttfamily 1405.5537}}].

\bibitem{Harman:2015gif}
C.~P.~D. Harman and S.~J. Huber, \emph{{Does zero temperature decide on the
  nature of the electroweak phase transition?}},
  \href{https://doi.org/10.1007/JHEP06(2016)005}{\emph{JHEP} {\bfseries 06}
  (2016) 005} [\href{https://arxiv.org/abs/1512.05611}{{\ttfamily
  1512.05611}}].

\bibitem{Basler:2016obg}
P.~Basler, M.~Krause, M.~Muhlleitner, J.~Wittbrodt and A.~Wlotzka,
  \emph{{Strong First Order Electroweak Phase Transition in the CP-Conserving
  2HDM Revisited}}, \href{https://doi.org/10.1007/JHEP02(2017)121}{\emph{JHEP}
  {\bfseries 02} (2017) 121}
  [\href{https://arxiv.org/abs/1612.04086}{{\ttfamily 1612.04086}}].

\bibitem{Dorsch:2017nza}
G.~C. Dorsch, S.~J. Huber, K.~Mimasu and J.~M. No, \emph{{The Higgs Vacuum
  Uplifted: Revisiting the Electroweak Phase Transition with a Second Higgs
  Doublet}}, \href{https://doi.org/10.1007/JHEP12(2017)086}{\emph{JHEP}
  {\bfseries 12} (2017) 086}
  [\href{https://arxiv.org/abs/1705.09186}{{\ttfamily 1705.09186}}].

\bibitem{Bernon:2017jgv}
J.~Bernon, L.~Bian and Y.~Jiang, \emph{{A new insight into the phase transition
  in the early Universe with two Higgs doublets}},
  \href{https://doi.org/10.1007/JHEP05(2018)151}{\emph{JHEP} {\bfseries 05}
  (2018) 151} [\href{https://arxiv.org/abs/1712.08430}{{\ttfamily
  1712.08430}}].

\bibitem{Andersen:2017ika}
J.~O. Andersen, T.~Gorda, A.~Helset, L.~Niemi, T.~V.~I. Tenkanen, A.~Tranberg
  et~al., \emph{{Nonperturbative Analysis of the Electroweak Phase Transition
  in the Two Higgs Doublet Model}},
  \href{https://doi.org/10.1103/PhysRevLett.121.191802}{\emph{Phys. Rev. Lett.}
  {\bfseries 121} (2018) 191802}
  [\href{https://arxiv.org/abs/1711.09849}{{\ttfamily 1711.09849}}].

\bibitem{Kainulainen:2019kyp}
K.~Kainulainen, V.~Keus, L.~Niemi, K.~Rummukainen, T.~V. Tenkanen and
  V.~Vaskonen, \emph{{On the validity of perturbative studies of the
  electroweak phase transition in the Two Higgs Doublet model}},
  \href{https://doi.org/10.1007/JHEP06(2019)075}{\emph{JHEP} {\bfseries 06}
  (2019) 075} [\href{https://arxiv.org/abs/1904.01329}{{\ttfamily
  1904.01329}}].

\bibitem{Zhou:2020xqi}
R.~Zhou and L.~Bian, \emph{{Baryon asymmetry and detectable Gravitational Waves
  from Electroweak phase transition}},
  \href{https://arxiv.org/abs/2001.01237}{{\ttfamily 2001.01237}}.

\bibitem{Carena:1996wj}
M.~Carena, M.~Quiros and C.~Wagner, \emph{{Opening the window for electroweak
  baryogenesis}},
  \href{https://doi.org/10.1016/0370-2693(96)00475-3}{\emph{Phys. Lett. B}
  {\bfseries 380} (1996) 81}
  [\href{https://arxiv.org/abs/hep-ph/9603420}{{\ttfamily hep-ph/9603420}}].

\bibitem{Delepine:1996vn}
D.~Delepine, J.~Gerard, R.~Gonzalez~Felipe and J.~Weyers, \emph{{A Light stop
  and electroweak baryogenesis}},
  \href{https://doi.org/10.1016/0370-2693(96)00921-5}{\emph{Phys. Lett. B}
  {\bfseries 386} (1996) 183}
  [\href{https://arxiv.org/abs/hep-ph/9604440}{{\ttfamily hep-ph/9604440}}].

\bibitem{Cline:1996cr}
J.~M. Cline and K.~Kainulainen, \emph{{Supersymmetric electroweak phase
  transition: Beyond perturbation theory}},
  \href{https://doi.org/10.1016/S0550-3213(96)00519-6}{\emph{Nucl. Phys. B}
  {\bfseries 482} (1996) 73}
  [\href{https://arxiv.org/abs/hep-ph/9605235}{{\ttfamily hep-ph/9605235}}].

\bibitem{Laine:1998qk}
M.~Laine and K.~Rummukainen, \emph{{The MSSM electroweak phase transition on
  the lattice}},
  \href{https://doi.org/10.1016/S0550-3213(98)00530-6}{\emph{Nucl. Phys. B}
  {\bfseries 535} (1998) 423}
  [\href{https://arxiv.org/abs/hep-lat/9804019}{{\ttfamily hep-lat/9804019}}].

\bibitem{Carena:2008vj}
M.~Carena, G.~Nardini, M.~Quiros and C.~Wagner, \emph{{The Baryogenesis Window
  in the MSSM}},
  \href{https://doi.org/10.1016/j.nuclphysb.2008.12.014}{\emph{Nucl. Phys. B}
  {\bfseries 812} (2009) 243}
  [\href{https://arxiv.org/abs/0809.3760}{{\ttfamily 0809.3760}}].

\bibitem{Cohen:2012zza}
T.~Cohen, D.~E. Morrissey and A.~Pierce, \emph{{Electroweak Baryogenesis and
  Higgs Signatures}},
  \href{https://doi.org/10.1103/PhysRevD.86.013009}{\emph{Phys. Rev. D}
  {\bfseries 86} (2012) 013009}
  [\href{https://arxiv.org/abs/1203.2924}{{\ttfamily 1203.2924}}].

\bibitem{Laine:2012jy}
M.~Laine, G.~Nardini and K.~Rummukainen, \emph{{Lattice study of an electroweak
  phase transition at m\_h \textasciitilde\ 126 GeV}},
  \href{https://doi.org/10.1088/1475-7516/2013/01/011}{\emph{JCAP} {\bfseries
  01} (2013) 011} [\href{https://arxiv.org/abs/1211.7344}{{\ttfamily
  1211.7344}}].

\bibitem{Curtin:2012aa}
D.~Curtin, P.~Jaiswal and P.~Meade, \emph{{Excluding Electroweak Baryogenesis
  in the MSSM}}, \href{https://doi.org/10.1007/JHEP08(2012)005}{\emph{JHEP}
  {\bfseries 08} (2012) 005} [\href{https://arxiv.org/abs/1203.2932}{{\ttfamily
  1203.2932}}].

\bibitem{Carena:2012np}
M.~Carena, G.~Nardini, M.~Quiros and C.~E. Wagner, \emph{{MSSM Electroweak
  Baryogenesis and LHC Data}},
  \href{https://doi.org/10.1007/JHEP02(2013)001}{\emph{JHEP} {\bfseries 02}
  (2013) 001} [\href{https://arxiv.org/abs/1207.6330}{{\ttfamily 1207.6330}}].

\bibitem{Katz:2015uja}
A.~Katz, M.~Perelstein, M.~J. Ramsey-Musolf and P.~Winslow,
  \emph{{Stop-Catalyzed Baryogenesis Beyond the MSSM}},
  \href{https://doi.org/10.1103/PhysRevD.92.095019}{\emph{Phys. Rev. D}
  {\bfseries 92} (2015) 095019}
  [\href{https://arxiv.org/abs/1509.02934}{{\ttfamily 1509.02934}}].

\bibitem{MJRMTripletPheno}
P.~Fileviez~Perez, H.~H. Patel, M.~Ramsey-Musolf and K.~Wang, \emph{{Triplet
  Scalars and Dark Matter at the LHC}},
  \href{https://doi.org/10.1103/PhysRevD.79.055024}{\emph{Phys. Rev.}
  {\bfseries D79} (2009) 055024}
  [\href{https://arxiv.org/abs/0811.3957}{{\ttfamily 0811.3957}}].

\bibitem{Chowdhury:2011ga}
T.~A. Chowdhury, M.~Nemevsek, G.~Senjanovic and Y.~Zhang, \emph{{Dark Matter as
  the Trigger of Strong Electroweak Phase Transition}},
  \href{https://doi.org/10.1088/1475-7516/2012/02/029}{\emph{JCAP} {\bfseries
  02} (2012) 029} [\href{https://arxiv.org/abs/1110.5334}{{\ttfamily
  1110.5334}}].

\bibitem{StepInto}
H.~H. Patel and M.~J. Ramsey-Musolf, \emph{{Stepping Into Electroweak Symmetry
  Breaking: Phase Transitions and Higgs Phenomenology}},
  \href{https://doi.org/10.1103/PhysRevD.88.035013}{\emph{Phys. Rev.}
  {\bfseries D88} (2013) 035013}
  [\href{https://arxiv.org/abs/1212.5652}{{\ttfamily 1212.5652}}].

\bibitem{MorriseyTwoStep}
N.~Blinov, J.~Kozaczuk, D.~E. Morrissey and C.~Tamarit, \emph{{Electroweak
  Baryogenesis from Exotic Electroweak Symmetry Breaking}},
  \href{https://doi.org/10.1103/PhysRevD.92.035012}{\emph{Phys. Rev.}
  {\bfseries D92} (2015) 035012}
  [\href{https://arxiv.org/abs/1504.05195}{{\ttfamily 1504.05195}}].

\bibitem{mjrm_triplet_lattice_1}
L.~Niemi, H.~H. Patel, M.~J. Ramsey-Musolf, T.~V. Tenkanen and D.~J. Weir,
  \emph{{Electroweak phase transition in the real triplet extension of the SM:
  Dimensional reduction}},
  \href{https://doi.org/10.1103/PhysRevD.100.035002}{\emph{Phys. Rev. D}
  {\bfseries 100} (2019) 035002}
  [\href{https://arxiv.org/abs/1802.10500}{{\ttfamily 1802.10500}}].

\bibitem{MJRMMultiplets}
W.~Chao, G.-J. Ding, X.-G. He and M.~Ramsey-Musolf, \emph{{Scalar Electroweak
  Multiplet Dark Matter}},  \href{https://arxiv.org/abs/1812.07829}{{\ttfamily
  1812.07829}}.

\bibitem{TripletPheno}
N.~F. Bell, M.~J. Dolan, L.~S. Friedrich, M.~J. Ramsey-Musolf and R.~R. Volkas,
  \emph{{Two-Step Electroweak Symmetry-Breaking: Theory Meets Experiment}},
  \href{https://doi.org/10.1007/JHEP05(2020)050}{\emph{JHEP} {\bfseries 20}
  (2020) 050} [\href{https://arxiv.org/abs/2001.05335}{{\ttfamily
  2001.05335}}].

\bibitem{100TeV_Triplet_Pheno}
C.-W. Chiang, G.~Cottin, Y.~Du, K.~Fuyuto and M.~J. Ramsey-Musolf,
  \emph{{Collider Probes of Real Triplet Scalar Dark Matter}},
  \href{https://arxiv.org/abs/2003.07867}{{\ttfamily 2003.07867}}.

\bibitem{mjrm_triplet_lattice_2}
L.~Niemi, M.~Ramsey-Musolf, T.~V. Tenkanen and D.~J. Weir,
  \emph{{Thermodynamics of a two-step electroweak phase transition}},
  \href{https://arxiv.org/abs/2005.11332}{{\ttfamily 2005.11332}}.

\bibitem{StrumiaMinimalDM}
M.~Cirelli, N.~Fornengo and A.~Strumia, \emph{{Minimal dark matter}},
  \href{https://doi.org/10.1016/j.nuclphysb.2006.07.012}{\emph{Nucl. Phys.}
  {\bfseries B753} (2006) 178}
  [\href{https://arxiv.org/abs/hep-ph/0512090}{{\ttfamily hep-ph/0512090}}].

\bibitem{StrumiaSommerfeld}
M.~Cirelli, A.~Strumia and M.~Tamburini, \emph{{Cosmology and Astrophysics of
  Minimal Dark Matter}},
  \href{https://doi.org/10.1016/j.nuclphysb.2007.07.023}{\emph{Nucl. Phys.}
  {\bfseries B787} (2007) 152}
  [\href{https://arxiv.org/abs/0706.4071}{{\ttfamily 0706.4071}}].

\bibitem{StrumiaCosmic}
M.~Cirelli, R.~Franceschini and A.~Strumia, \emph{{Minimal Dark Matter
  predictions for galactic positrons, anti-protons, photons}},
  \href{https://doi.org/10.1016/j.nuclphysb.2008.03.013}{\emph{Nucl. Phys.}
  {\bfseries B800} (2008) 204}
  [\href{https://arxiv.org/abs/0802.3378}{{\ttfamily 0802.3378}}].

\bibitem{TripletDM2}
O.~{Fischer} and J.~J. {van der Bij}, \emph{{Multi-Singlet and Singlet-Triplet
  Scalar Dark Matter}},
  \href{https://doi.org/10.1142/S0217732311036528}{\emph{Modern Physics Letters
  A} {\bfseries 26} (2011) 2039}.

\bibitem{MultipletEWPTDM}
S.~S. AbdusSalam and T.~A. Chowdhury, \emph{{Scalar Representations in the
  Light of Electroweak Phase Transition and Cold Dark Matter Phenomenology}},
  \href{https://doi.org/10.1088/1475-7516/2014/05/026}{\emph{JCAP} {\bfseries
  1405} (2014) 026} [\href{https://arxiv.org/abs/1310.8152}{{\ttfamily
  1310.8152}}].

\bibitem{TripletDM1}
O.~Fischer and J.~J. van~der Bij, \emph{{The scalar Singlet-Triplet Dark Matter
  Model}}, \href{https://doi.org/10.1088/1475-7516/2014/01/032}{\emph{JCAP}
  {\bfseries 1401} (2014) 032}
  [\href{https://arxiv.org/abs/1311.1077}{{\ttfamily 1311.1077}}].

\bibitem{LHCTripDM}
S.~Yaser~Ayazi and S.~M. Firouzabadi, \emph{{Constraining Inert Triplet Dark
  Matter by the LHC and FermiLAT}},
  \href{https://doi.org/10.1088/1475-7516/2014/11/005}{\emph{JCAP} {\bfseries
  1411} (2014) 005} [\href{https://arxiv.org/abs/1408.0654}{{\ttfamily
  1408.0654}}].

\bibitem{TripDMFootprint}
S.~Yaser~Ayazi and S.~M. Firouzabadi, \emph{{Footprint of Triplet Scalar Dark
  Matter in Direct, Indirect Search and Invisible Higgs Decay}},
  \href{https://doi.org/10.1080/23311940.2015.1047559}{\emph{Cogent Phys.}
  {\bfseries 2} (2015) 1047559}
  [\href{https://arxiv.org/abs/1501.06176}{{\ttfamily 1501.06176}}].

\bibitem{TripletGravWaves}
M.~Chala, M.~Ramos and M.~Spannowsky, \emph{{Gravitational wave and collider
  probes of a triplet Higgs sector with a low cutoff}},
  \href{https://doi.org/10.1140/epjc/s10052-019-6655-1}{\emph{Eur. Phys. J.}
  {\bfseries C79} (2019) 156}
  [\href{https://arxiv.org/abs/1812.01901}{{\ttfamily 1812.01901}}].

\bibitem{LHCTripletPheno2013}
L.~Wang and X.-F. Han, \emph{{LHC diphoton and Z+photon Higgs signals in the
  Higgs triplet model with Y = 0}},
  \href{https://doi.org/10.1007/JHEP03(2014)010}{\emph{JHEP} {\bfseries 03}
  (2014) 010} [\href{https://arxiv.org/abs/1303.4490}{{\ttfamily 1303.4490}}].

\bibitem{planckTriplet}
N.~Khan, \emph{{Exploring the hyperchargeless Higgs triplet model up to the
  Planck scale}},
  \href{https://doi.org/10.1140/epjc/s10052-018-5766-4}{\emph{Eur. Phys. J.}
  {\bfseries C78} (2018) 341}
  [\href{https://arxiv.org/abs/1610.03178}{{\ttfamily 1610.03178}}].

\bibitem{LHCTripletPheno}
M.~Chabab, M.~C. Peyranère and L.~Rahili, \emph{{Probing the Higgs sector of
  $Y=0$ Higgs Triplet Model at LHC}},
  \href{https://doi.org/10.1140/epjc/s10052-018-6339-2}{\emph{Eur. Phys. J.}
  {\bfseries C78} (2018) 873}
  [\href{https://arxiv.org/abs/1805.00286}{{\ttfamily 1805.00286}}].

\bibitem{Georgi:1985nv}
H.~Georgi and M.~Machacek, \emph{{DOUBLY CHARGED HIGGS BOSONS}},
  \href{https://doi.org/10.1016/0550-3213(85)90325-6}{\emph{Nucl. Phys. B}
  {\bfseries 262} (1985) 463}.

\bibitem{Chanowitz:1985ug}
M.~S. Chanowitz and M.~Golden, \emph{{Higgs Boson Triplets With M ($W$) = M
  ($Z$) $\cos \theta \omega$}},
  \href{https://doi.org/10.1016/0370-2693(85)90700-2}{\emph{Phys. Lett. B}
  {\bfseries 165} (1985) 105}.

\bibitem{Zhou:2018zli}
R.~Zhou, W.~Cheng, X.~Deng, L.~Bian and Y.~Wu, \emph{{Electroweak phase
  transition and Higgs phenomenology in the Georgi-Machacek model}},
  \href{https://doi.org/10.1007/JHEP01(2019)216}{\emph{JHEP} {\bfseries 01}
  (2019) 216} [\href{https://arxiv.org/abs/1812.06217}{{\ttfamily
  1812.06217}}].

\bibitem{Zhou:2020idp}
L.~Bian, H.-K. Guo, Y.~Wu and R.~Zhou, \emph{{Gravitational wave and collider
  searches for electroweak symmetry breaking patterns}},
  \href{https://doi.org/10.1103/PhysRevD.101.035011}{\emph{Phys. Rev. D}
  {\bfseries 101} (2020) 035011}
  [\href{https://arxiv.org/abs/1906.11664}{{\ttfamily 1906.11664}}].

\bibitem{THNMSSM}
P.~Bandyopadhyay, C.~Coriano and A.~Costantini, \emph{{Perspectives on a
  supersymmetric extension of the standard model with a Y = 0 Higgs triplet and
  a singlet at the LHC}},
  \href{https://doi.org/10.1007/JHEP09(2015)045}{\emph{JHEP} {\bfseries 09}
  (2015) 045} [\href{https://arxiv.org/abs/1506.03634}{{\ttfamily
  1506.03634}}].

\bibitem{SO10SUSYTriplet}
S.~A.~R. Ellis, T.~Gherghetta, K.~Kaneta and K.~A. Olive, \emph{{New Weak-Scale
  Physics from SO(10) with High-Scale Supersymmetry}},
  \href{https://doi.org/10.1103/PhysRevD.98.055009}{\emph{Phys. Rev.}
  {\bfseries D98} (2018) 055009}
  [\href{https://arxiv.org/abs/1807.06488}{{\ttfamily 1807.06488}}].

\bibitem{TwoStep}
S.~Inoue, G.~Ovanesyan and M.~J. Ramsey-Musolf, \emph{{Two-Step Electroweak
  Baryogenesis}}, \href{https://doi.org/10.1103/PhysRevD.93.015013}{\emph{Phys.
  Rev.} {\bfseries D93} (2016) 015013}
  [\href{https://arxiv.org/abs/1508.05404}{{\ttfamily 1508.05404}}].

\bibitem{Two_component_SigmaxSM}
A.~Dutta~Banik, R.~Roshan and A.~Sil, \emph{{Two Component Singlet-Triplet
  Scalar Dark Matter and Electroweak Vacuum Stability}},
  \href{https://arxiv.org/abs/2009.01262}{{\ttfamily 2009.01262}}.

\bibitem{gambit_xsm_dm_scalar}
{\scshape GAMBIT} collaboration, \emph{{Status of the scalar singlet dark
  matter model}},
  \href{https://doi.org/10.1140/epjc/s10052-017-5113-1}{\emph{Eur. Phys. J. C}
  {\bfseries 77} (2017) 568}
  [\href{https://arxiv.org/abs/1705.07931}{{\ttfamily 1705.07931}}].

\bibitem{higgs_portal_dm_review}
G.~Arcadi, A.~Djouadi and M.~Raidal, \emph{{Dark Matter through the Higgs
  portal}}, \href{https://doi.org/10.1016/j.physrep.2019.11.003}{\emph{Phys.
  Rept.} {\bfseries 842} (2020) 1}
  [\href{https://arxiv.org/abs/1903.03616}{{\ttfamily 1903.03616}}].

\bibitem{pdg2020}
{\scshape Particle Data Group} collaboration, \emph{{Review of Particle
  Physics}}, \href{https://doi.org/10.1093/ptep/ptaa104}{\emph{PTEP} {\bfseries
  2020} (2020) 083C01}.

\bibitem{Forshaw:2001xq}
J.~R. Forshaw, D.~Ross and B.~White, \emph{{Higgs mass bounds in a triplet
  model}}, \href{https://doi.org/10.1088/1126-6708/2001/10/007}{\emph{JHEP}
  {\bfseries 10} (2001) 007}
  [\href{https://arxiv.org/abs/hep-ph/0107232}{{\ttfamily hep-ph/0107232}}].

\bibitem{bounded_general_potential}
K.~Kannike, \emph{{Vacuum Stability of a General Scalar Potential of a Few
  Fields}}, \href{https://doi.org/10.1140/epjc/s10052-016-4160-3}{\emph{Eur.
  Phys. J. C} {\bfseries 76} (2016) 324}
  [\href{https://arxiv.org/abs/1603.02680}{{\ttfamily 1603.02680}}].

\bibitem{mjrmRunningCouplingsSinglets}
M.~Gonderinger, H.~Lim and M.~J. Ramsey-Musolf, \emph{{Complex Scalar Singlet
  Dark Matter: Vacuum Stability and Phenomenology}},
  \href{https://doi.org/10.1103/PhysRevD.86.043511}{\emph{Phys. Rev.}
  {\bfseries D86} (2012) 043511}
  [\href{https://arxiv.org/abs/1202.1316}{{\ttfamily 1202.1316}}].

\bibitem{mjrmRunningCouplings}
Y.~Du, A.~Dunbrack, M.~J. Ramsey-Musolf and J.-H. Yu, \emph{{Type-II Seesaw
  Scalar Triplet Model at a 100 TeV $pp$ Collider: Discovery and Higgs Portal
  Coupling Determination}},
  \href{https://doi.org/10.1007/JHEP01(2019)101}{\emph{JHEP} {\bfseries 01}
  (2019) 101} [\href{https://arxiv.org/abs/1810.09450}{{\ttfamily
  1810.09450}}].

\bibitem{micromegas}
G.~Bélanger, F.~Boudjema, A.~Goudelis, A.~Pukhov and B.~Zaldivar,
  \emph{{micrOMEGAs5.0 : Freeze-in}},
  \href{https://doi.org/10.1016/j.cpc.2018.04.027}{\emph{Comput. Phys. Commun.}
  {\bfseries 231} (2018) 173}
  [\href{https://arxiv.org/abs/1801.03509}{{\ttfamily 1801.03509}}].

\bibitem{planck2018}
{\scshape Planck} collaboration, \emph{{Planck 2018 results. VI. Cosmological
  parameters}},  \href{https://arxiv.org/abs/1807.06209}{{\ttfamily
  1807.06209}}.

\bibitem{XENON1T1Y}
{\scshape XENON} collaboration, \emph{{Dark Matter Search Results from a One
  Ton-Year Exposure of XENON1T}},
  \href{https://doi.org/10.1103/PhysRevLett.121.111302}{\emph{Phys. Rev. Lett.}
  {\bfseries 121} (2018) 111302}
  [\href{https://arxiv.org/abs/1805.12562}{{\ttfamily 1805.12562}}].

\bibitem{Ackermann:2015lka}
{\scshape Fermi-LAT} collaboration, \emph{{Updated search for spectral lines
  from Galactic dark matter interactions with pass 8 data from the Fermi Large
  Area Telescope}},
  \href{https://doi.org/10.1103/PhysRevD.91.122002}{\emph{Phys. Rev. D}
  {\bfseries 91} (2015) 122002}
  [\href{https://arxiv.org/abs/1506.00013}{{\ttfamily 1506.00013}}].

\bibitem{madgraph}
J.~Alwall, R.~Frederix, S.~Frixione, V.~Hirschi, F.~Maltoni, O.~Mattelaer
  et~al., \emph{{The automated computation of tree-level and next-to-leading
  order differential cross sections, and their matching to parton shower
  simulations}}, \href{https://doi.org/10.1007/JHEP07(2014)079}{\emph{JHEP}
  {\bfseries 07} (2014) 079} [\href{https://arxiv.org/abs/1405.0301}{{\ttfamily
  1405.0301}}].

\bibitem{prospino_chargino}
W.~Beenakker, M.~Klasen, M.~Kramer, T.~Plehn, M.~Spira and P.~Zerwas,
  \emph{{The Production of charginos / neutralinos and sleptons at hadron
  colliders}},
  \href{https://doi.org/10.1103/PhysRevLett.100.029901}{\emph{Phys. Rev. Lett.}
  {\bfseries 83} (1999) 3780}
  [\href{https://arxiv.org/abs/hep-ph/9906298}{{\ttfamily hep-ph/9906298}}].

\bibitem{madwidth}
J.~Alwall, C.~Duhr, B.~Fuks, O.~Mattelaer, D.~G. Öztürk and C.-H. Shen,
  \emph{{Computing decay rates for new physics theories with FeynRules and
  MadGraph 5\_aMC@NLO}},
  \href{https://doi.org/10.1016/j.cpc.2015.08.031}{\emph{Comput. Phys. Commun.}
  {\bfseries 197} (2015) 312}
  [\href{https://arxiv.org/abs/1402.1178}{{\ttfamily 1402.1178}}].

\bibitem{cms_chargino_neutralino_higgs_36fb}
{\scshape CMS} collaboration, \emph{{Combined search for electroweak production
  of charginos and neutralinos in proton-proton collisions at $\sqrt{s} =$ 13
  TeV}}, \href{https://doi.org/10.1007/JHEP03(2018)160}{\emph{JHEP} {\bfseries
  03} (2018) 160} [\href{https://arxiv.org/abs/1801.03957}{{\ttfamily
  1801.03957}}].

\bibitem{atlas_chargino_neutralino_higgs_36fb}
{\scshape ATLAS} collaboration, \emph{{Search for chargino and neutralino
  production in final states with a Higgs boson and missing transverse momentum
  at $\sqrt{s} = 13$ TeV with the ATLAS detector}},
  \href{https://doi.org/10.1103/PhysRevD.100.012006}{\emph{Phys. Rev. D}
  {\bfseries 100} (2019) 012006}
  [\href{https://arxiv.org/abs/1812.09432}{{\ttfamily 1812.09432}}].

\bibitem{atlas_chargino_neutralino_compressed_140fb}
{\scshape ATLAS} collaboration, \emph{{Searches for electroweak production of
  supersymmetric particles with compressed mass spectra in $\sqrt{s}=$ 13 TeV
  $pp$ collisions with the ATLAS detector}},
  \href{https://doi.org/10.1103/PhysRevD.101.052005}{\emph{Phys. Rev. D}
  {\bfseries 101} (2020) 052005}
  [\href{https://arxiv.org/abs/1911.12606}{{\ttfamily 1911.12606}}].

\bibitem{cms_chargino_neutralino_compressed_36fb}
{\scshape CMS} collaboration, \emph{{Search for supersymmetry with a compressed
  mass spectrum in the vector boson fusion topology with 1-lepton and 0-lepton
  final states in proton-proton collisions at $\sqrt{s}=$ 13 TeV}},
  \href{https://doi.org/10.1007/JHEP08(2019)150}{\emph{JHEP} {\bfseries 08}
  (2019) 150} [\href{https://arxiv.org/abs/1905.13059}{{\ttfamily
  1905.13059}}].

\bibitem{checkmate}
D.~Dercks, N.~Desai, J.~S. Kim, K.~Rolbiecki, J.~Tattersall and T.~Weber,
  \emph{{CheckMATE 2: From the model to the limit}},
  \href{https://doi.org/10.1016/j.cpc.2017.08.021}{\emph{Comput. Phys. Commun.}
  {\bfseries 221} (2017) 383}
  [\href{https://arxiv.org/abs/1611.09856}{{\ttfamily 1611.09856}}].

\bibitem{cms_chargino_neutralino_diphoton_77fb}
{\scshape CMS} collaboration, \emph{{Search for supersymmetry using Higgs boson
  to diphoton decays at $\sqrt{s}$ = 13 TeV}},
  \href{https://doi.org/10.1007/JHEP11(2019)109}{\emph{JHEP} {\bfseries 11}
  (2019) 109} [\href{https://arxiv.org/abs/1908.08500}{{\ttfamily
  1908.08500}}].

\bibitem{pythia3}
T.~Sjöstrand, S.~Ask, J.~R. Christiansen, R.~Corke, N.~Desai, P.~Ilten et~al.,
  \emph{{An Introduction to PYTHIA 8.2}},
  \href{https://doi.org/10.1016/j.cpc.2015.01.024}{\emph{Comput. Phys. Commun.}
  {\bfseries 191} (2015) 159}
  [\href{https://arxiv.org/abs/1410.3012}{{\ttfamily 1410.3012}}].

\bibitem{MadAnalysis5}
E.~Conte and B.~Fuks, \emph{{Confronting new physics theories to LHC data with
  MADANALYSIS 5}}, \href{https://doi.org/10.1142/S0217751X18300272}{\emph{Int.
  J. Mod. Phys. A} {\bfseries 33} (2018) 1830027}
  [\href{https://arxiv.org/abs/1808.00480}{{\ttfamily 1808.00480}}].

\bibitem{atlas_chargino_neutralino_compressed_140fb_f41ab}
{\scshape ATLAS} collaboration, \emph{{'Figure 41ab' of 'Searches for
  electroweak production of supersymmetric particles with compressed mass
  spectra in $\sqrt{s}=13$ TeV $pp$ collisions with the ATLAS detector'}},
  2020.
\newblock 10.17182/hepdata.91374.v2/t79.

\bibitem{atlas_chargino_neutralino_compressed_140fb_f41cd}
{\scshape ATLAS} collaboration, \emph{{'Figure 41cd' of 'Searches for
  electroweak production of supersymmetric particles with compressed mass
  spectra in $\sqrt{s}=13$ TeV $pp$ collisions with the ATLAS detector'}},
  2020.
\newblock 10.17182/hepdata.91374.v2/t80.

\bibitem{atlas_chargino_neutralino_diphoton_140fb}
{\scshape ATLAS} collaboration, \emph{{Search for direct production of
  electroweakinos in final states with missing transverse momentum and a Higgs
  boson decaying into photons in $pp$ collisions at $\sqrt{s}=13$ TeV with the
  ATLAS detector}},  \href{https://arxiv.org/abs/2004.10894}{{\ttfamily
  2004.10894}}.

\bibitem{atlas_chargino_neutralino_Z_140fb}
{\scshape ATLAS} collaboration, \emph{{Search for chargino-neutralino
  production with mass splittings near the electroweak scale in three-lepton
  final states in $\sqrt {s}$=13 TeV $pp$ collisions with the ATLAS detector}},
  \href{https://doi.org/10.1103/PhysRevD.101.072001}{\emph{Phys. Rev. D}
  {\bfseries 101} (2020) 072001}
  [\href{https://arxiv.org/abs/1912.08479}{{\ttfamily 1912.08479}}].

\bibitem{atlas_chargino_neutralino_ww_140fb}
{\scshape ATLAS} collaboration, \emph{{Search for electroweak production of
  charginos and sleptons decaying into final states with two leptons and
  missing transverse momentum in $\sqrt{s}=13$ TeV $pp$ collisions using the
  ATLAS detector}},
  \href{https://doi.org/10.1140/epjc/s10052-019-7594-6}{\emph{Eur. Phys. J. C}
  {\bfseries 80} (2020) 123}
  [\href{https://arxiv.org/abs/1908.08215}{{\ttfamily 1908.08215}}].

\bibitem{atlas_chargino_neutralino_ww_140fb_t45}
{\scshape ATLAS} collaboration, \emph{{'xsec upper limits 1' of 'Search for
  electroweak production of charginos and sleptons decaying into final states
  with two leptons and missing transverse momentum in $\sqrt{s}=13$ TeV $pp$
  collisions using the ATLAS detector'}},  2019.
\newblock 10.17182/hepdata.89413.v1/t45.

\bibitem{Profumo:2010kp}
S.~Profumo, L.~Ubaldi and C.~Wainwright, \emph{{Singlet Scalar Dark Matter:
  monochromatic gamma rays and metastable vacua}},
  \href{https://doi.org/10.1103/PhysRevD.82.123514}{\emph{Phys. Rev. D}
  {\bfseries 82} (2010) 123514}
  [\href{https://arxiv.org/abs/1009.5377}{{\ttfamily 1009.5377}}].

\bibitem{Duerr:2015aka}
M.~Duerr, P.~Fileviez~P\'erez and J.~Smirnov, \emph{{Scalar Dark Matter: Direct
  vs. Indirect Detection}},
  \href{https://doi.org/10.1007/JHEP06(2016)152}{\emph{JHEP} {\bfseries 06}
  (2016) 152} [\href{https://arxiv.org/abs/1509.04282}{{\ttfamily
  1509.04282}}].

\end{thebibliography}\endgroup

\end{document}